\documentclass[pdftex,11pt,a4paper]{article} 

\usepackage{authblk}
\usepackage[utf8]{inputenc}
\usepackage[round]{natbib}
\usepackage[pdftex]{graphicx}
\usepackage{amsmath,amsthm,amsfonts}
\usepackage{lscape}   
\usepackage{rotating}
\usepackage{rotfloat}
\usepackage{multirow}
\usepackage{booktabs}
\usepackage{ctable}
\usepackage{threeparttable}
\usepackage{booktabs}
\usepackage{framed}
\usepackage{color}
\usepackage[colorlinks=true,linkcolor=blue,allcolors=blue]{hyperref}

\usepackage{soul}
\newtheorem{theorem}{Theorem}[section] 
\newtheorem{corollary}[theorem]{Corollary} 
\newtheorem{definition}{Definition}[section] 
\usepackage[margin=10pt,font=small,labelfont=bf,
labelsep=endash]{caption} [2023/03/12]\usepackage[margin=10pt,font=small,labelfont=bf,
labelsep=endash]{caption} [2023/03/12]

\title{A Fuzzy Approach for Randomized Confidence Intervals}

\author{Carlos Henrique Trigo Nasser Felix}
\author{Nancy Lopes Garcia}
\author{Alex Rodrigo dos S. Sousa \thanks{asousa@unicamp.br}}

\affil{Universidade Estadual de Campinas (UNICAMP)\\ Departament of Statistics, Brazil}

\date{} 

\begin{document}

\maketitle

\vspace{-1.0cm}
\begin{abstract}
  We propose randomized confidence intervals based on the Neyman-Pearson lemma, in order to make them more broadly applicable to distributions that do not satisfy regularity conditions. This is achieved by using the definition of fuzzy confidence intervals. These intervals are compared with methods described in the literature for well-known distributions such as normal, binomial, and Poisson. The results show that in high-variance situations, the new intervals provide better performance. Furthermore, through these intervals, it is possible to compute a lower bound for the expected length, demonstrating that they achieve the minimal maximum expected length for a Bernoulli trial observation.
\\
  
\noindent{\bf Keywords:} Confidence intervals. Neyman–Pearson lemma. Measure theory. Expected length. Fuzzy logic. \\
  \end{abstract}

\section{Introduction}

The construction of confidence intervals dates back to \cite{LaPlace1820}, who developed a confidence interval for the parameter representing the proportion of successes in a binomial distribution. The proposed interval is centered at a point estimator, the sample proportion, and has a length proportional to the standard error of the estimator. However, this construction can present difficulties when applied for parameters of a discrete distribution, as its coverage probability may not match the desired confidence level. An approach that has been widely adopted is to allow the coverage probability to exceed the reference value. As shown in \cite{Agresti1998}, this approach generates wider intervals compared to asymptotic methods, for which the coverage rate approaches the confidence level as the sample size increases. Therefore, asymptotic confidence intervals are suitable if the sample size is large enough so that the coverage rate is not significantly lower than the confidence level.

One way to overcome this problem and achieve a nominal coverage rate is to use randomized confidence intervals as initially proposed in \cite{Stevens1950}. Later \cite{Geyer2005} proposed a method to construct fuzzy confidence intervals based on the uniformly most powerful unbiased hypothesis test and the definition of a fuzzy logic membership function. A comparison of this method with others in the literature can be found in \cite{tese}. See also \cite{fuzzy} for a general exposition of fuzzy theory.

The present work aims to develop a fuzzy confidence interval methodology based on the Neyman-Pearson lemma for simple hypotheses, designed to be applicable to any parametric family of distributions while achieving a smaller expected length at a specific value compared to any other method. In addition, the proposed estimator was constructed and compared with other methods present in the literature for the normal with bounded parameter space, binomial, and Poisson distributions.

An interesting application of the proposed method arises from the classical knapsack problem. In this context, the knapsack problem can be naturally linked to fuzzy confidence intervals by interpreting the selection of intervals as an optimization process under uncertainty and imprecision. Each candidate interval is regarded as an item whose “value” represents its confidence level or coverage, while its “weight” corresponds to the interval length or an associated expected cost. Fuzzy membership functions are employed to capture the gradual satisfaction of confidence requirements, rather than imposing a strict binary inclusion rule. Consequently, the resulting optimization problem aims to balance the maximization of overall confidence against the minimization of interval width, closely mirroring the objective of selecting an optimal subset of items in a knapsack subject to capacity constraints. See \cite{knapsack} for an overview of the knapsack problem.

This work is organized as follows. Section 2 presents the main results, while their proofs are provided in Section 4. The knapsack problem is developed in Section 3. Section 5 illustrates applications of the proposed methodology to the binomial, Poisson, and normal distributions. Final remarks and concluding considerations are given in Section 6.

The proposed method is implemented and available in the R package \texttt{FRCI} of \cite{package}.

\section{Main Results}

To understand the proposed method, it is essential to recall the well-known duality between confidence intervals and hypothesis testing. In classical statistical inference, a confidence interval for a parameter can be interpreted as the set of parameter values that are not rejected by a corresponding family of hypothesis tests at a given significance level. Conversely, a hypothesis test can be derived by checking whether a hypothesized parameter value lies within the confidence interval.

This duality is particularly transparent in the Neyman–Pearson framework, where hypothesis tests are constructed to control type I error probabilities. Confidence intervals arise naturally by inverting these tests: for each candidate parameter value, a test is performed, and the collection of values for which the null hypothesis is not rejected forms the confidence set. When regularity conditions hold, this inversion leads to standard, non-randomized confidence intervals with exact or asymptotic coverage.

However, in settings involving discrete distributions or non-regular models, the direct inversion of non-randomized tests may fail to achieve the nominal coverage level. In such cases, randomized tests play a crucial role, and their inversion leads to randomized—or fuzzy—confidence intervals. The proposed method exploits this testing–interval duality by constructing fuzzy confidence intervals through optimal tests derived from the Neyman–Pearson lemma, thereby ensuring correct coverage while allowing for improved efficiency, as measured by expected interval length.

Let $\Theta$ denote a parametric space, and consider $(\Omega,\mathcal A,\mu(\cdot \mid \theta))$ a probability space indexed by $\theta \in \Theta$ and denote $\mathbb{E}_{\theta}$ the corresponding expectation functional. 

\begin{definition}\label{def:1}[Fuzzy Confidence Interval] Let $\gamma \in (0,1)$ be the desired confidence level. An $\mathcal{A}$-measurable membership function
\[
\psi : \Omega \times \Theta \to [0,1]
\]
satisfying
\[
\int_{\Omega} \psi(\omega \mid \tau)\, d\mu(\omega \mid \tau) \ge \gamma
\]
defines a fuzzy confidence interval with $100\gamma\%$ confidence.
\end{definition}

\begin{definition}[Randomized Neyman-Pearson Test] Let $1-\gamma \in (0,1)$ be the significance level of the test  $H_0: \theta=\tau$. Then the randomized rejection region is given by the rejection function $\psi(\tau, \cdot): \Omega \rightarrow [0,1]$. 
\end{definition}

\begin{definition}[Expected Fuzzy Length] Let $\gamma \in (0,1)$ be the desired confidence level. Let $(\Theta, \mathcal{O}, \nu)$ be a measure space defining the size on $\Theta$, and let $\psi$ be the membership function defining a fuzzy confidence interval. The quantity
\[
\mathrm{EL}(\psi, \theta, \nu)
= \mathbb{E}_{\theta}\!\left[\int_{\Theta} \psi(\omega \mid \tau)\, d\nu(\tau)\right]
\]
is called the expected length of the fuzzy confidence interval.
\end{definition}

\subsection{Minimization problems}

The approach consists of selecting a reference value $o \in \Theta$ and, given this choice, constructing a function 
$$\psi_o:\Omega\times\Theta \to [0,1]$$
that minimizes the probability of failing to reject $H_0: \theta= o$ when the distribution associated with $o$ is taken as the true one, for any $\tau \neq o$.

For $\psi_o$ to define a confidence interval, as required in Definition \ref{def:1}, the probability of failing to reject $\tau$ when the distribution associated with $\tau$ is the true one be greater than or equal to $\gamma$, for all $\tau \in \Theta$. This condition serves as the constraint of the minimization problem.

The constraint can be expressed directly on the set over which the functions 
$\psi$ are evaluated. Specifically, we aim to minimize 
\begin{equation}
   \int_\Omega \psi(\omega \mid \tau)\, d\mu(\omega \mid o), \nonumber
\end{equation}
for $\psi \in \mathfrak{F}_\gamma$, where
\begin{eqnarray*}
\mathfrak{F_\gamma} &=& \left\{\psi:\Omega\times\Theta\to [0,1]\ \Bigg|\  \psi(\cdot \mid \tau)\ \text{is}\ \mathcal{A}/\mathcal{B}(\mathbb{R})\text{-measurable and } \right. \\
&&\hspace{5cm} \left.\int_\Omega\psi(\omega \mid \tau)d\mu(\omega \mid \tau)\ge \gamma , \forall\tau\in\Theta \right\}.
\end{eqnarray*}

First, we will present the nonparametric version of the theorem of interest. The parametric version follows as a corollary.

\begin{theorem}\label{thm:1} Let $(\Omega,\mathcal{A},\mu)$ and $(\Omega,\mathcal{A},\nu)$ be two probability spaces, and let $\gamma\in(0,1)$ denote the desired confidence level. Define
\begin{eqnarray*}
\mathfrak{E}(\gamma,\mu) &=& \left\{\psi:\Omega\to [0,1]\ \Bigg|\  \psi\ \text{is}\ \mathcal{A}\text{-measurable and } \int_\Omega \psi(\omega)\, d\mu(\omega)\ge \gamma \right\}
\end{eqnarray*}
as the set of $\mathcal{A}$-measurable functions with coverage probability greater than $\gamma$. Then there exists $\psi^*\in\mathfrak{E}(\gamma,\mu)$ such that, for every $\psi\in\mathfrak{E}(\gamma,\mu)$,
$$
\int_\Omega \psi(\omega)\, d\nu(\omega)
\ge
\int_\Omega \psi^*(\omega)\, d\nu(\omega).
$$
\end{theorem}

To construct the randomized confidence interval, we use some interpretations of Theorem \ref{thm:1}, which can be seen as a hypothesis test with simple hypotheses, with $H_0: \mu$ describing the distribution of the data versus $H_1: \nu$ describing the distribution of the data. We can parametrize these measures in a parametric family while maintaining the property of being the Uniformly Most Powerful (UMP) test.

Consider the parametric space $\Theta$ and the values $o$ and $\tau$ $\in \Theta$, writing the associated probability measures as $\mu(\cdot \mid o)$ and $\mu(\cdot \mid \tau)$, respectively. We can obtain the membership function given $o$ by combining the functions indexed by $\tau$ defined as the minimizing function $\psi_\tau^*:\Omega\to[0,1]$ with $\psi_\tau^*\in \mathfrak{E}(\gamma,\mu(\cdot \mid \tau))$, $\mu=\mu(\cdot \mid \tau)$, and $\nu=\mu(\cdot \mid o)$, defining $\psi_o(\omega \mid \tau)=\psi_\tau^*(\omega)$. Therefore, $\psi_o\in\mathfrak{F_\gamma}$ and $\psi_o$ is UMP for $H_0: \theta=o$.

\begin{corollary}\label{thm:2} Let $\Theta$ be a parametric space, $(\Omega,\mathcal{A},\mu(\cdot \mid \tau))$ a probability space for all $\tau\in\Theta$, and $\gamma\in(0,1)$ the desired confidence level. Let 
\begin{eqnarray*}
\mathfrak{F_\gamma} &=& \left\{\psi:\Omega\times\Theta\to [0,1]\ \Bigg|\  \psi(\cdot \mid \tau)\ \text{is}\ \mathcal{A}/\mathcal{B}(\mathbb{R})\text{-measurable and } \right. \\
&&\hspace{5cm} \left.\int_\Omega\psi(\omega \mid \tau)d\mu(\omega|\tau)\ge \gamma , \forall\tau\in\Theta \right\}
\end{eqnarray*}
be the set of $\mathcal{A}$-measurable functions with coverage rate at least $\gamma$. Then, for a given $o\in\Theta$, there exists $\psi^*\in\mathfrak{F_\gamma}$ such that for all $\psi\in\mathfrak{F_\gamma}$,
$$\int_\Omega\psi(\omega \mid \tau)d\mu(\omega \mid o)\ge\int_\Omega\psi^* (\tau|\omega)d\mu(\omega \mid o).$$
\end{corollary}

This theorem requires fewer conditions than the Neyman-Pearson Lemma, but in exchange, the uniqueness property is lost. In the analogous case of the ``knapsack problem", the switch from a confidence interval problem to a fuzzy confidence interval problem allows one to express the solution in terms of the evaluation of the Radon-Nikodym derivative, making it simpler.

\subsection{Expected Interval Length}
Given a measure space $(\Theta,\mathcal{O},\nu)$, 
the expected length of a function $\psi(\omega \mid \tau)$ with respect to the measure $\nu$ is given by 
$\mathrm{EL}(\psi,\theta,\nu)=\int_\Theta\int_\Omega\psi(\omega \mid \tau)d\mu(\omega|\theta)d\nu(\tau)$, 
if $\int_\Omega\psi(\omega \mid \tau)d\mu(\omega|\theta)$ is $\mathcal{O}$-measurable.

\begin{theorem}\label{thm:3} Given the conditions and definitions in Theorem \ref{thm:1}, consider the measure space $(\Theta,\mathcal{O},\nu)$ such that, for every $o\in\Theta$,
$$(\theta,\tau)\mapsto\int_\Omega\psi_{o}(\omega \mid \tau)d\mu(\omega,\theta)$$
is $\mathcal{O}$-measurable. Let 
the subset $\mathfrak{M_\gamma}$ of $\mathfrak{F_\gamma}$ be defined by 
\begin{equation*}
    \mathfrak{M_\gamma}=\left\{\psi\in\mathfrak{F_\gamma}\ \middle|\ (\theta,\tau)\mapsto\int_\Omega\psi(\omega \mid \tau)d\mu(\omega,\theta)\mbox{ is $\mathcal{O}$-measurable}\right\}
\end{equation*}

and define the function $\mathrm{EL}(\theta,\psi):\Theta\times\mathfrak{M_\gamma}\to[0,+\infty]$ by 
$$\mathrm{EL}(\theta,\psi)=\int_\Theta\int_\Omega\psi(\omega \mid \tau)d\mu(\omega,\theta)d\nu(\tau),$$
then for every $\gamma\in(0,1)$ and $\theta\in\Theta$ there exists $\psi^*\in\mathfrak{M_\gamma}$ such that
\begin{equation*}
    \mathrm{EL}(\theta,\psi)\ge \mathrm{EL}(\theta,\psi^*), \mbox{ for all $\psi\in\mathfrak{M_\gamma}$.}
\end{equation*}
$$$$
\end{theorem}

\section{The knapsack problem}

\ \ The knapsack problem has several formulations in the literature. In this text, we will consider the following formulation: consider a collection of $n$ objects numbered from $1$ to $n$, where object $i$ has weight and value given by $w_i$ and $v_i$ respectively, with $i=1,...,n$, and a knapsack capable of carrying a subset of these objects, such that the sum of the weights of the selected objects does not exceed the maximum weight limit of $W$. The goal is to find a subset of objects that has the largest sum of values such that it does not exceed the maximum weight capacity of the knapsack $W$.

Consider $x_i\in\{0,1\}$ as a variable indicating whether object $i$ is in the knapsack. The problem can be formulated as minimizing $\sum_{i=1}^nv_ix_i$ restricted to $\sum_{i=1}^nw_ix_i\le W$, whose exact solution can be obtained by calculating all possible combinations. However, depending on the weights, values, and total number of objects, this may be unfeasible, and in this case, dynamic programming can be used.

\begin{figure}[H]
\centering
\graphicspath{{img/}}
\includegraphics[width=5cm]{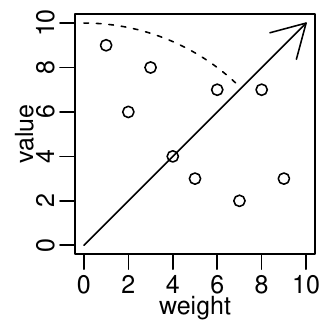}
\caption{Example of the graphical solution described in George B. Dantzig (1957) for the knapsack problem.}\label{fig:diag}
\end{figure}

If the condition $x_i\in\{0,1\}$ is relaxed to $x_i\in[0,1]$, the solution can be described simply. Furthermore, this new condition can be interpreted as the possibility of selecting only a part of the object, such that the value and weight are proportional to the totality of the selected object, see \cite{Dantzig1957}. An example of a solution can be seen in Figure \ref{fig:diag}, which contains the scatter plot of the value versus the weight of the objects. The half-line divides the objects into 3 groups: those above the segment, those below, and those contained within, which will be called $A$, $B$, and $C$ respectively. The solution is given by placing all the points above the segment and part of those contained within the segments in the knapsack, such that the total weight of the knapsack is equal to the weight limit.

Start with the half-line originating at $(0,0)$ and initially parallel to the \textbf{value} axis, meaning all objects are in $B$. Then rotate the half-line clockwise until it contains at least one object and its intersection with the half-line before rotation is only the origin. The objects contained in the half-line are in set $C$. Now evaluate if the sum of the weights of the objects in $A\cup C$ is greater than $W$. If not, continue rotating the half-line, causing the objects in $C$ to move to $A$. If so, stop the process.

With the sets $A$, $B$, and $C$, the solution can be obtained by adding the objects from $A$ to the knapsack; those in $B$ will not be added to the knapsack; the objects in $C$ will be partially added, with the proportion $\frac{W-W_A}{W_C}$, with $W_A$ being the sum of the weights of the objects in $A$ and $W_C$ being the sum of the weights of the objects in $C$, we can write the solution as follows.

\begin{equation}\label{eq:PmXDef}
x_i = \left\{ \begin{array}{lcl}
1 &,\ \mathrm{if}& i\in A \\
0 &,\ \mathrm{if} & i\in B \\
\frac{W-W_A}{W_C} &,\ \mathrm{if} & i\in C, 
\end{array}\right.
\end{equation}
for $i=1,2,...,n.$

The connection between Theorem~1 and the knapsack problem becomes apparent when the theorem is viewed as a special case with 
\[
\Omega = \{1,2,\ldots,n\}, \quad 
\gamma = 1 - \frac{W}{\sum_{i=1}^n w_i}, \quad 
\mu(\{i\}) = \frac{w_i}{\sum_{i=1}^n w_i}, \quad 
\nu(\{i\}) = \frac{v_i}{\sum_{i=1}^n v_i}.
\]
Under this formulation, the optimal solution satisfies $w_i^{*} = 1 - x_i$. Moreover, the strategy of partitioning the elements into fully kept, discarded, and partially kept sets plays a central role in the proof of Theorem~1.

\section{Proofs}
In this section we provide the proofs of Theorems \ref{thm:1} and \ref{thm:3}. 

\subsection{Proof of Theorem \ref{thm:1}}

  The main part of this proof is based on the partition of the sample space $\Omega$ into four disjoint sets similarly as the randomized version of the knapsack problem. The increase in the number of partitions is due to the need to include a term to account for the singular part of the Lebesgue decomposition of $\nu$ with respect to $\mu$. The sets $D$ and $D^c$ are defined such that there exist measures $\nu_1$ and $\nu_2$ satisfying, for $O\in \Omega$, $\nu_1(O)=\nu(D\cap O)$ and $\nu_2(O)=\nu(D^c\cap O)$, with $\nu_1\perp\mu$ and $\nu_2\ll\mu$.

\ \ \ \ \ \ According to the Radon-Nikodym Theorem, since $\nu_2\ll\mu$, there exists a function $\frac{d\nu_2}{d\mu}:\Omega\to[0,+\infty]$ such that
$$\displaystyle\nu_2(O)=\int_O\frac{d\nu_2}{d\mu}(\omega)d\mu(\omega),$$
and as $\frac{d\nu_2}{d\mu}$ is a measurable function on a measure space, we can view it as a random variable $Y$. Using random variable notation, $Y(\omega)=\frac{d\nu_2}{d\mu}(\omega)\ge0$.

With this notation, we define the cumulative distribution function $F(y)=P[Y\le y]$ and its inverse, the quantile function $Q:[0,1]\to[0,\infty)$, defined by $Q(p)=\inf\{x\in \mathbb{R}:p\le F(x)\}$.

Based on this, given $\gamma\in(0,1)$, we define the following sets
\begin{eqnarray*}
  A_\gamma&=&[Y<Q(\gamma)], \\
  B_\gamma&=&[Y>Q(\gamma)], \\
  C_\gamma&=&[Y=Q(\gamma)].
\end{eqnarray*}

Note that $A_\gamma,B_\gamma,C_\gamma,D \in \mathcal{A}$, $A_\gamma\cup B_\gamma\cup C_\gamma \cup D = \Omega$, $\mu(A_\gamma)\le\gamma$ and $\mu(B_\gamma)\le 1-\gamma$. 

Given $\gamma\in(0,1)$, we define the function $\psi^*:\Omega\to[0,1]$ as follows

\begin{equation}
\label{eq:DefPsiS}
\psi^*(\omega)=
\left\{
\begin{array}{ccl}
1 &,\ \mathrm{if}& \omega\in A_\gamma \\
0 &,\ \mathrm{if}& \omega\in B_\gamma\cup D_\gamma \\
\frac{\gamma-\mu(A_\gamma)}{\mu(C_\gamma)} &,\ \mathrm{if}& \omega\in C_\gamma\mbox{ and } \mu(C_\gamma)\neq0\\
0&\ \mathrm{if}&\omega\in C_\gamma\mbox{ and } \mu(C_\gamma)=0.
\end{array}
\right. \nonumber
\end{equation}

Note that $\psi^*\in \mathfrak{E}(\gamma,\mu)$ since $\psi^*$ is $\mathcal{A}$-measurable as it is a simple $\mathcal{A}$-function, and $\int_\Omega \psi^*(\omega)d\mu(\omega)\ge \gamma$. Thus,

\begin{eqnarray}
\label{eq:IntPsiS}
    \lefteqn{\int_\Omega \psi^*(\omega)d\mu(\omega)} \nonumber \\
    & & =\int_{A_\gamma}\psi^*(\omega)d\mu(\omega)+\int_{B_\gamma}\psi^*(\omega)d\mu(\omega)+\int_{C_\gamma}\psi^*(\omega)d\mu(\omega)\nonumber \\
    & & =\int_{A_\gamma}d\mu(\omega)+0+\int_{C_\gamma}\frac{\gamma-\mu(A_\gamma)}{\mu(C_\gamma)}d\mu(\omega) \nonumber \\
    & & =\mu(A_\gamma)+\frac{\gamma-\mu(A_\gamma)}{\mu(C_\gamma)}(\mu(C_\gamma))\nonumber \\
    & & =\mu(A_\gamma)+(\gamma-\mu(A_\gamma))\nonumber \\
    & & =\gamma. \nonumber
\end{eqnarray}

Now we will show that for all $\psi\in\mathfrak{E}(\gamma,\mu)$,
$$\int_\Omega\psi(\omega)d\nu(\omega)\ge\int_\Omega\psi^*(\omega)d\nu(\omega),$$
i.e., we will show that $\int_\Omega\psi(\omega)-\psi^*(\omega)d\nu(\omega)\ge 0$.

First, we split $\Omega$ into $A_\gamma,B_\gamma,C_\gamma\ \text{and}\ D$ in the integral. Since $\psi^*(\omega)=0$ for $\omega\in D$, we have $\psi(\omega)-\psi^*(\omega)=\psi(\omega)\ge0$, and thus the value of the integral over $\Omega$ is greater than the value over $A_{\gamma}\cup B_{\gamma}\cup C_{\gamma}=D^c$. In this case,

\begin{equation}
\begin{array}{rcl}
    \int_\Omega\psi(\omega)-\psi^*(\omega)d\nu(\omega)& \ge & \int_{A_\gamma}\psi(\omega)-\psi^*(\omega)d\nu(\omega), \\
     &&  \quad + \int_{B_\gamma}\psi(\omega)-\psi^*(\omega)d\nu(\omega), \\
     && \quad + \int_{C_\gamma}\psi(\omega)-\psi^*(\omega)d\nu(\omega). 
\end{array} \nonumber
\end{equation}

Using the Radon-Nikodym derivative property $X(\omega)$, we can rewrite an integral over a set contained in $D^c$ with respect to $\nu$ as an integral over the same set with respect to $\mu$ as follows:

\begin{eqnarray*}
     \int_{A_\gamma}\psi(\omega)-\psi^*(\omega)d\nu(\omega)  &=& \int_{A_\gamma}X(\omega)(\psi(\omega)-\psi^*(\omega))d\mu(\omega), \\
      \int_{B_\gamma}\psi(\omega)-\psi^*(\omega)d\nu(\omega) &=& \int_{B_\gamma}X(\omega)(\psi(\omega)-\psi^*(\omega))d\mu(\omega), \\
     \int_{C_\gamma}\psi(\omega)-\psi^*(\omega)d\nu(\omega) &=& \int_{C_\gamma}X(\omega)(\psi(\omega)-\psi^*(\omega))d\mu(\omega).
\end{eqnarray*}

Note that

\begin{itemize}
    \item for $\omega\in A_\gamma$ we have $\psi^*(\omega)=1\ge\psi(\omega)$ and therefore $\psi(\omega)-\psi^*(\omega)\le 0$,
    \item for $\omega\in B_\gamma$ we have $\psi^*(\omega)=0\le\psi(\omega)$ and therefore $\psi(\omega)-\psi^*(\omega)\ge 0$. 
\end{itemize} 

Furthermore, by the definition of the sets $A_\gamma=[Y<Q(\gamma)]$, $B_\gamma=[Y>Q(\gamma)]$ and $C_\gamma=[Y=Q(\gamma)]$, we have

 \begin{eqnarray}
      \int_{A_\gamma}Y(\omega)(\psi(\omega)-\psi^*(\omega))d\mu(\omega) & \ge & Q(\gamma) \int_{A_\gamma}(\psi(\omega)-\psi^*(\omega))d\mu(\omega), \label{eq:t1} \\
       \int_{B_\gamma}Y(\omega)(\psi(\omega)-\psi^*(\omega))d\mu(\omega) &\ge& Q(\gamma) \int_{B_\gamma}(\psi(\omega)-\psi^*(\omega))d\mu(\omega), \label{eq:t2} \\
      \int_{C_\gamma}Y(\omega)(\psi(\omega)-\psi^*(\omega)d\mu(\omega)) &=& Q(\gamma) \int_{C_\gamma}(\psi(\omega)-\psi^*(\omega)d\mu(\omega)). \label{eq:t3} 
 \end{eqnarray}

Note that by the definition of $D$ and $D^c$ we have $\mu(D)=0$, and therefore summing equations (\ref{eq:t1})--(\ref{eq:t3}) we obtain

\begin{equation}
    \int_{D^c}X(\omega)(\psi(\omega)-\psi^*(\omega))d\mu(\omega)\ge Q(\gamma)\int_{\Omega}(\psi(\omega)-\psi^*(\omega))d\mu(\omega). \nonumber
\end{equation}

To conclude the proof, it is enough to show that
$Q(\gamma)\int_{\Omega}(\psi(\omega)-\psi^*(\omega))d\mu(\omega)\ge 0$. 

Since $Q(\gamma)\ge 0$ and $\psi\in\mathfrak{E}(\gamma,\mu)$ its cover tax is greater than the confidence level $\gamma$, while  $\psi^* = \gamma$, therefore $\int_\Omega\psi(\omega)d\mu-\int_\Omega\psi^*(\omega)d\mu(\omega)=\int_\Omega\psi(\omega)d\mu-\gamma\ge 0$.

\subsection{Proof of Theorem \ref{thm:3}}

    From the previous proof, we have that 
$\int_\Omega \psi(\omega \mid  \tau) - \psi_\theta(\omega \mid \tau) \, d\mu(\omega,\theta) \ge 0$, hence
\begin{eqnarray*}
\lefteqn{\int_\Theta \int_\Omega \psi(\omega \mid  \tau) - \psi_\theta(\omega \mid \tau) \, d\mu(\omega \mid \theta) \, d\nu(\tau) \ge 0} \\
&\iff& \int_\Theta \int_\Omega \psi(\omega \mid  \tau) \, d\mu(\omega \mid \theta) \, d\nu(\tau) \ge \int_\Theta \int_\Omega \psi_\theta(\omega \mid \tau) \, d\mu(\omega \mid \theta) \, d\nu(\tau) \\
&\iff& \mathrm{EL}(\theta, \psi) \ge \mathrm{EL}(\theta, \psi_\theta).
\end{eqnarray*}

\section{Examples}
In this section we provide examples of the application of Theorems \ref{thm:1} and \ref{thm:3} and Corollary \ref{thm:2} in the binomial, Poisson and normal distributions.

\subsection{Binomial distribution}

\subsubsection{Fuzzy pertinent function}

 Consider the case of a random sample of size 1 with a binomial distribution with parameters $n$ and $\theta \in \Theta = (0,1)$. In this case, we have $\Omega = {0,1,2,\ldots,n}$, $\mathcal{A} = \mathcal{P}(\Omega)$, and the counting measure $\#:\mathcal{P}(\Omega)\to[0,\infty]$ and 
 $$\displaystyle \mu(A \mid \theta)=\int_A {\binom n \omega} \theta^\omega(1-\theta)^{n-\omega} 1_\Omega(\omega)d\#(\omega)=\sum_{\omega\in A}{ \binom{n}{\omega} } \theta^\omega(1-\theta)^{n-\omega}.$$

Since ${\binom n \omega} \theta^\omega(1-\theta)^{n-\omega}>0$ for all $\theta>0$ and $\omega\in \Omega$, we have that $D^c=\Omega$ and we can obtain
\begin{eqnarray*}
Y(\omega)&=&
\frac{d\mu(\cdot \mid o)}{d\mu(\cdot \mid \tau)}(\omega) \\
&=&
\left(\frac{o}{1-o}\right)^\omega(1-o)^n\left(\frac{1-\tau}{\tau}\right)^\omega(1-\tau)^{-n}
\\
&=&
\left(\frac{o}{1-o}\frac{1-\tau}{\tau}\right)^\omega\left(\frac{1-o}{1-\tau}\right)^n.
\end{eqnarray*}

To define the sets $A_\gamma$, $B_\gamma$, and $C_\gamma$, it is possible to compute the quantile function of $Y$ at $\gamma$, denoted by $Q(\gamma)$. However, it is easier to find an equivalence to the sets $[Y < Q(\gamma)]$, $[Y = Q(\gamma)]$, and $[Y > Q(\gamma)]$ by using an auxiliary random variable.

Defining the random variable by the function $X(\omega)=\omega$, we have $X\sim \mathrm{Bin} (n,\tau)$ and $Y(\omega)=\left(\frac{o}{1-o}\frac{1-\tau}{\tau}\right)^{X(\omega)}\left(\frac{1-o}{1-\tau}\right)^n$, for $o\neq \tau$ and $o,\tau\in(0,1)$, we can rewrite it as $X(\omega)=\frac{\ln(Y(\omega))-n\ln\left(\frac{1-o}{1-\tau}\right)}{\ln\left(\frac{o}{1-o}\frac{1-\tau}{\tau}\right)}$, note that the function is increasing in $Y(\omega)$ for $o>\tau$ and decreasing if $o<\tau$, then we will divide it into two cases and write the sets $A_\gamma$, $B_\gamma$, $C_\gamma$ as a function of $X$.

\paragraph*{Case 1: $\tau<o$}

\begin{itemize}
\item 
$\begin{aligned}[t] A_\gamma &=[Y<Q(\gamma)]=
\left[\frac{\ln(Y(\omega))-n\ln\left(\frac{1-o}{1-\tau}\right)}{\ln\left(\frac{o}{1-o}\frac{1-\tau}{\tau}\right)}<\frac{\ln(Q(\gamma))-n\ln\left(\frac{1-o}{1-\tau}\right)}{\ln\left(\frac{o}{1-o}\frac{1-\tau}{\tau}\right)}\right]\\
&=\left[X<\frac{\ln(Q(\gamma))-n\ln\left(\frac{1-o}{1-\tau}\right)}{\ln\left(\frac{o}{1-o}\frac{1-\tau}{\tau}\right)}\right],
\end{aligned}$
\item 
$\begin{aligned}[t] B_\gamma &=[Y>Q(\gamma)]=
\left[\frac{\ln(Y(\omega))-n\ln\left(\frac{1-o}{1-\tau}\right)}{\ln\left(\frac{o}{1-o}\frac{1-\tau}{\tau}\right)}>\frac{\ln(Q(\gamma))-n\ln\left(\frac{1-o}{1-\tau}\right)}{\ln\left(\frac{o}{1-o}\frac{1-\tau}{\tau}\right)}\right],\\
&=\left[X>\frac{\ln(Q(\gamma))-n\ln\left(\frac{1-o}{1-\tau}\right)}{\ln\left(\frac{o}{1-o}\frac{1-\tau}{\tau}\right)}\right].
\end{aligned}$

\item 
$\begin{aligned}[t] C_\gamma &=[Y=Q(\gamma)]=\left[\frac{\ln(Y(\omega))-n\ln\left(\frac{1-o}{1-\tau}\right)}{\ln\left(\frac{o}{1-o}\frac{1-\tau}{\tau}\right)}=\frac{\ln(Q(\gamma))-n\ln\left(\frac{1-o}{1-\tau}\right)}{\ln\left(\frac{o}{1-o}\frac{1-\tau}{\tau}\right)}\right]\\
&=\left[X=\frac{\ln(Q(\gamma))-n\ln\left(\frac{1-o}{1-\tau}\right)}{\ln\left(\frac{o}{1-o}\frac{1-\tau}{\tau}\right)}\right]
\end{aligned}$

\end{itemize}

Since $X\sim \mathrm{Bin}(n,\tau)$, we have that $Q_X(\gamma)\in\{0,...,n\}$ e $\mu(\{0,...,Q_X(\gamma)-1\}|\tau)<\gamma$  and $\mu(\{0,...,Q_X(\gamma)\}|\tau)\ge\gamma$. By the definitions of $A_\gamma$ and $B_\gamma$, we have

\begin{itemize}
    \item $P\left[X<\frac{\ln(Q(\gamma))-n\ln\left(\frac{1-o}{1-\tau}\right)}{\ln\left(\frac{o}{1-o}\frac{1-\tau}{\tau}\right)}\right]\le\gamma,$
    \item $P\left[X>\frac{\ln(Q(\gamma))-n\ln\left(\frac{1-o}{1-\tau}\right)}{\ln\left(\frac{o}{1-o}\frac{1-\tau}{\tau}\right)}\right]\le1-\gamma.$
\end{itemize}

Then, we can define the sets as $A_\gamma=[X<Q_X(\gamma)]$, $B_\gamma=[X>Q_X(\gamma)]$ e $C_\gamma=[X=Q_X(\gamma)]$.

 Using the relationship between the cumulative probability of the binomial distribution and the regularized beta function defined by $I(x,a,b)=\frac{\int_0^xt^{a-1}(1-t)^{b-1}dt}{\int_0^1t^{a-1}(1-t)^{b-1}dt}$, we have

\begin{itemize}
    \item $\mu(\{0,...,Q_X(\gamma)-1\}|\tau)=1-I(\tau,Q_X(\gamma),n-Q_X(\gamma)+1)<\gamma$, e
    \item $\mu(\{0,...,Q_X(\gamma)\}|\tau)=1-I(\tau,Q_X(\gamma)+1,n-Q_X(\gamma))\ge\gamma$.
\end{itemize}

These inequalities are equivalent to

\begin{itemize}
    \item $\tau> I^{-1}(1-\gamma,Q_X(\gamma),n-Q_X(\gamma)+1)$, e
    \item $\tau\le I^{-1}(1-\gamma,Q_X(\gamma)+1,n-Q_X(\gamma))$.
\end{itemize}

\noindent Since the right-hand side of the inequalities can be written as a function of $I^{-1}(1-\gamma,i,n-i+1)$, with $i=Q_X(\gamma)$ for the upper part and $i=Q_X(\gamma)+1$ for the lower part, because it is an increasing function as a function of $i$, we can define $Q_X(\gamma)$ as follows:\\
$Q_X(\gamma)=\sup \{i\in\{0,1,2,...,n\}|\tau> I^{-1}(1-\gamma,i,n-i+1)\}$.

\begin{figure}[H]
\graphicspath{{img/}}
\centering
\includegraphics[width=15cm]{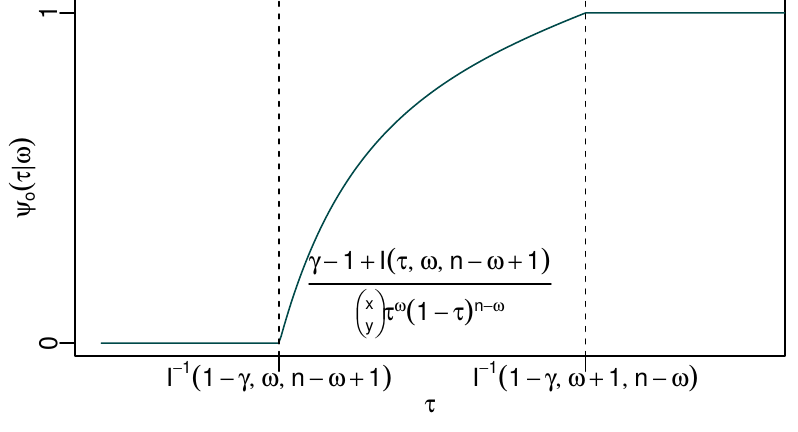}
\caption{Fuzzy membership function of the method developed for the value of $\tau$ for the binomial case with $\tau<o$.}
\label{fig:binom1}
\end{figure}

In other words, we have that $Q_X(\gamma)=i$ if, and only if, $I^{-1}(1-\gamma,i,n-i+1)<\tau\le I^{-1}(1-\gamma,i+1,n-i)$. In this condition we have the definitions of the sets $A_\gamma=\{0,1,...,i-1\}$, $B_\gamma=\{i+1,i+2,...,n\}$ and $C_\gamma=\{i\}$, and therefore $\mu(A_\gamma|\tau)=1-I(\tau,i,n-i+1)$ and $\mu(C_\gamma|\tau)=\binom n \omega \tau^\omega(1-\tau)^{n-\omega}$, replacing the value of $i$ with $\omega$ it is possible to write the function $\psi_o(\omega \mid \tau)$ for the case $\tau<o$ present in \eqref{eq:binomor}. This function, which is depicted in Figure \ref{fig:binom1}, is non-decreasing with respect to $\tau$ with $\omega$ fixed, since it is the case $\tau<o$, the fuzzy membership function of the described method is non-decreasing in $\tau$ up to the value $o$. 

\begin{equation}
\label{eq:binomor}
     \psi_o(\omega \mid \tau) = \left\{
    \begin{array}{lcl}
        1&\mbox{, if}&
        \tau> I^{-1}(1-\gamma,\omega+1,n-\omega),
        \\
        0&\mbox{, if}&
        \tau\le I^{-1}(1-\gamma,\omega,n-\omega+1),
        \\
        \frac{\gamma-1+I(\tau,\omega,n-\omega+1)}{\binom n \omega \tau^\omega(1-\tau)^{n-\omega}}
        &\mbox{, if}&
        \left\{
        \begin{array}{l}
             \tau> I^{-1}(1-\gamma,\omega,n-\omega+1),\\
             \tau\le I^{-1}(1-\gamma,\omega+1,n-\omega).
        \end{array}
        \right.
    \end{array} \right. 
\end{equation}

\begin{figure}[H]
\graphicspath{{img/}}
\includegraphics[width=13cm]{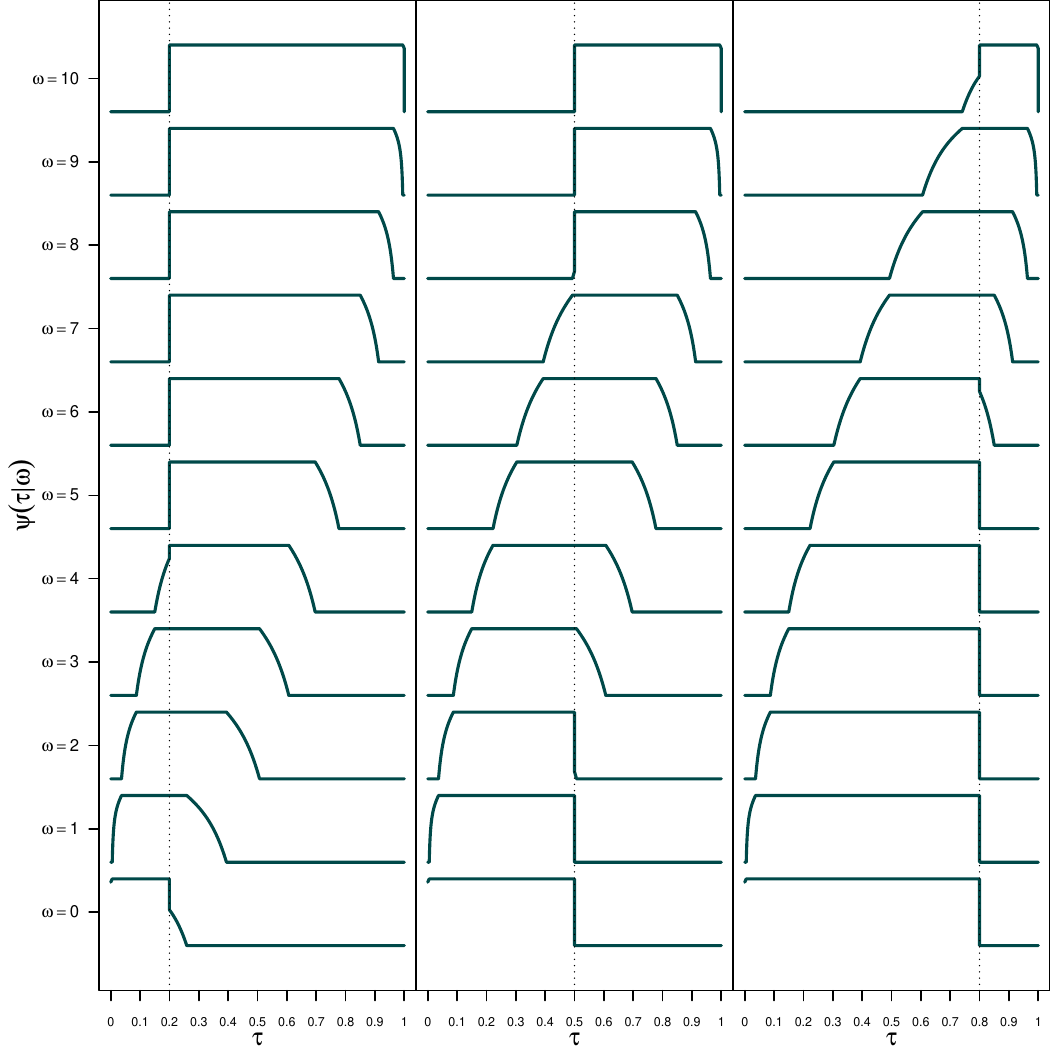}
\centering
\caption{Fuzzy membership function $\psi_o(\omega \mid \tau)$ in the case of the binomial distribution with $n=10$ and $o=0.2$ for the left panel, $o=0.5$ for the middle panel and $o=0.8$ in the right panel.} \label{fig:binom2}
\end{figure}

Similar for $\tau > o$.  Combining the two cases, the function $\psi_o$ is given by \eqref{eq:psiobinom}. The function is non-decreasing up to $o$ and non-increasing after that value when evaluated with respect to $\tau$ with $\omega$ and $o$ fixed. Figure \ref{fig:binom2} shows the Fuzzy membership function \eqref{eq:psiobinom} for $n = 10$ and $o = 0.2$, $0.5$ and $0.8$. The main characteristic of this function is that its left-hand limit or right-hand limit at $o$ is equal to $1$, with a discontinuity at $\omega$ far from the value $\lfloor o(n+1)\rfloor$, containing only values less than $o$ if $\omega<\lfloor o(n+1)\rfloor$, and only values greater than $o$ if $\omega>\lfloor o(n+1)\rfloor$. 

\begin{figure}[H]
\graphicspath{{img/}}
\includegraphics[width=15cm]{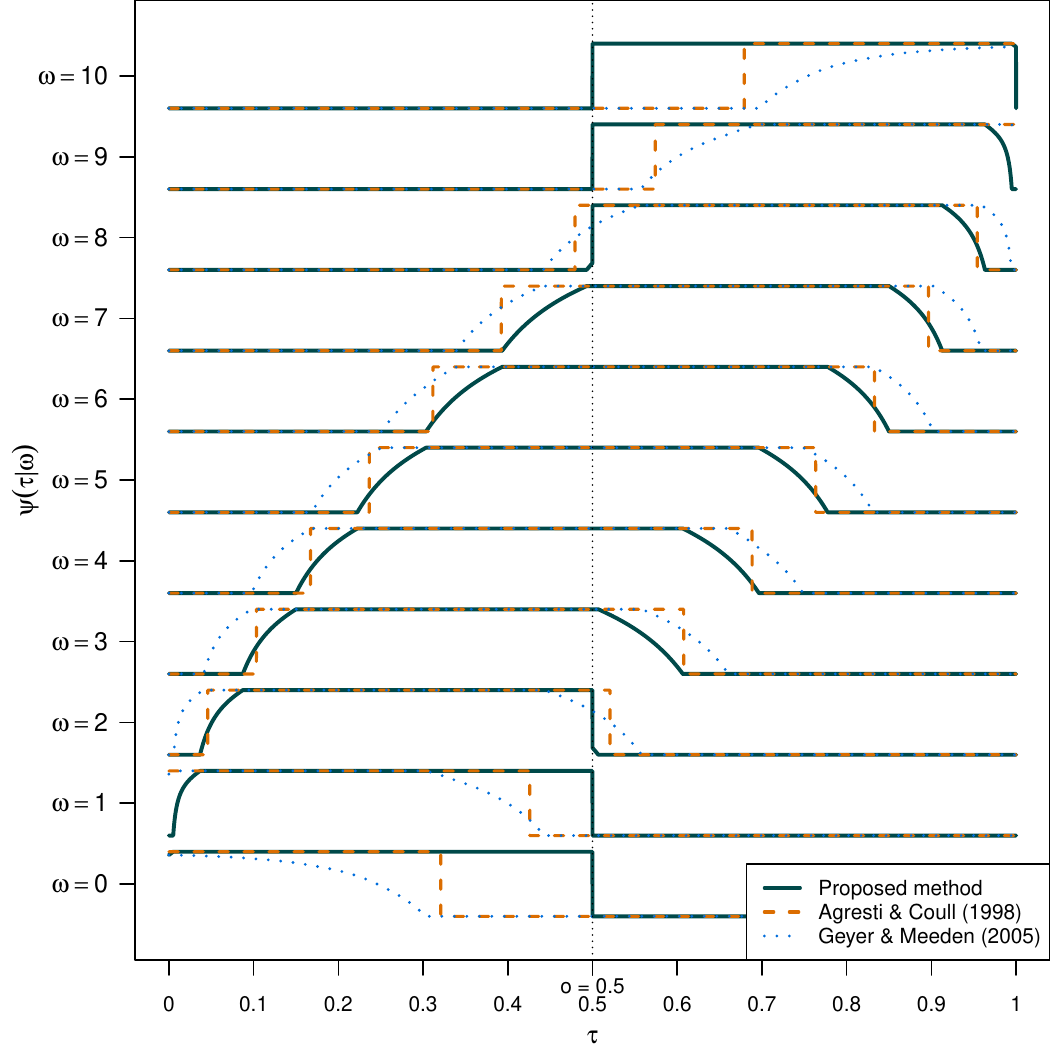}
\caption{Fuzzy membership functions $\psi_{0.5}$, $\psi^{AC}$ and $\psi^{GM}$ with confidence level $\gamma = 0.95$ for the binomial distribution with $n = 10$.} \label{fig:binom3}
\centering
\end{figure}

Finally, Figure \ref{fig:binom3} provides the Fuzzy membership functions for $n = 10$ and $\gamma = 0.95$ by the proposed method \eqref{eq:psiobinom} for $o = 0.5$ and by the Agresti-Coull and Geyer-Meeden methods.


\begin{equation}
\label{eq:psiobinom}
    \psi_o(\omega \mid \tau) = \left\{
    \begin{array}{rcl}
        1&\mbox{, if}&
            o < \tau\le I^{-1}(\gamma,\omega,n-\omega+1),\\
        1&\mbox{, if}&
             I^{-1}(1-\gamma,\omega+1,n-\omega) < \tau < o,\\
        0&\mbox{, if}&
            \tau> \max(I^{-1}(\gamma,\omega+1,n-\omega),o)\\
        0&\mbox{, if}&
            \tau\le I^{-1}(1-\gamma,\omega,n-\omega+1), \mbox{ and } 
            \tau<o,\\
        \frac{\gamma-I(\tau,\omega+1,n-\omega)}{\binom n \omega \tau^\omega(1-\tau)^{n-\omega}},
        &\mbox{, if}&
        \left\{
        \begin{array}{l}
             \tau>\max(I^{-1}(\gamma,\omega,n-\omega+1),0),\\
             \tau\le I^{-1}(\gamma,\omega+1,n-\omega),
        \end{array}
        \right.
        \\
        \frac{\gamma-1+I(\tau,\omega,n-\omega+1)}{\binom n \omega \tau^\omega(1-\tau)^{n-\omega}},
         &\mbox{, if}&
         \left\{
         \begin{array}{l}
            \tau> I^{-1}(1-\gamma,\omega,n-\omega+1),\\
            \tau\le I^{-1}(1-\gamma,\omega+1,n-\omega), \mbox{ and } \tau<o.\\        
         \end{array}
         \right.
    \end{array} \right.
\end{equation}

\subsubsection{Expected interval length}
\begin{figure}[H]
\graphicspath{{img/}}
\includegraphics[width=15cm]{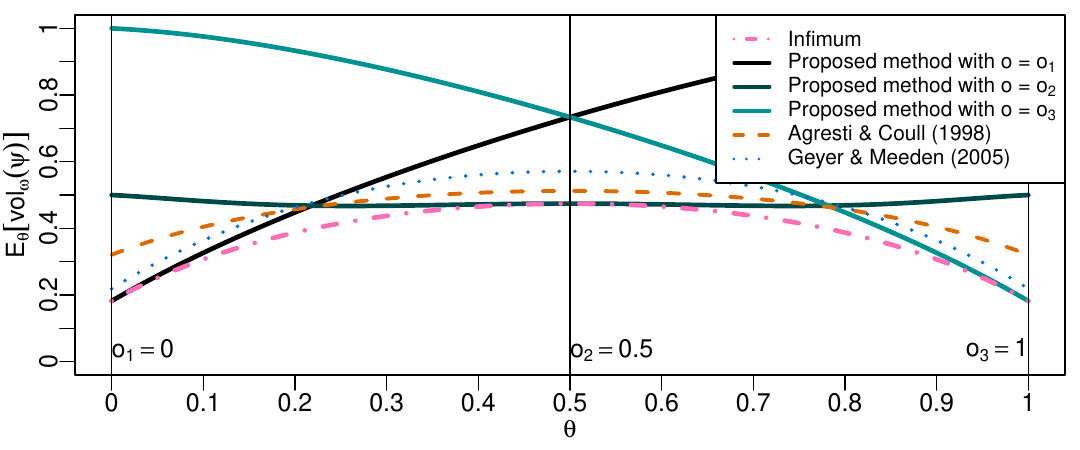}
\caption{Expected interval length in the binomial case with parameter $n = 10$ and $\theta$, for the five methods discussed, three of which correspond to $\psi_o$ for $o_1 = 0.1$, $o_2 = 0.5$, and $o_3 = 0.9$, and the methods $\psi^{GM}$ and $\psi^{AC}$. The black dotted curve shows the lower bound obtained by $TE(\theta, \psi_\theta, \lambda)$, where $\lambda$ refers to the Lebesgue measure.} \label{fig:binom4}
\centering
\end{figure}

In this example, we present the calculation of the expected interval length for the proportion $\theta$ of the binomial distribution using the proposed method for $o \in {0, 0.5, 1}$ and the methods of Geyer–Meeden and Agresti–Coull. It is important to note that, in this case, the calculations were carried out numerically in order to obtain Figure \ref{fig:binom4} for $n = 10$.

\subsection{Poisson distribution}

\subsubsection{Fuzzy pertinent function}
Consider the case of an observation with a Poisson distribution with parameter $\theta\in\Theta=(0,\infty)$. In this case, we have $\Omega=\mathbb{N} ,\ \mathcal{A}=\mathcal{P}(\mathbb{N})$, the counting measure $\#:\mathcal{P}(\mathbb{N})\to[0,\infty]$ and $$\displaystyle \mu(A|\theta)=\int_A\frac{e^{-\theta}\theta^\omega}{\omega!}1_\Omega(\omega)d\#(\omega)=\sum_{\omega\in A}\frac{e^{-\theta}\theta^\omega}{\omega!}.$$

Analogously to the binomial distribution example, we have that
\begin{equation}
Y(\omega)=
e^{-(o-\tau)}\left(\frac{o}{\tau}\right)^\omega. \nonumber
\end{equation}

Defining the random variable by the function $X(\omega)=\omega$, we have $X\sim Poisson\left(\tau\right)$ and $Y(\omega)=e^{-(o-\tau)}\left(\frac{o}{\tau}\right)^{X(\omega)}$, for $o\neq \tau$ and $o,\tau>0$, we can rewrite it as $X(\omega)=\frac{o-\tau+ \ln(Y(\omega))}{\ln o - \ln \tau}$, note that the function is increasing in $Y(\omega)$ for $o>\tau$ and decreasing if $o<\tau$, so we will divide it into two cases and write the sets $A_\gamma$, $B_\gamma$, $C_\gamma$ as a function of $X$.

\paragraph*{Case 1: $\tau<o$}

\begin{itemize}
\item 
$\begin{aligned}[t] A_\gamma &=[Y<Q(\gamma)]=
\left[\frac{o-\tau+ \ln(Y(\omega))}{\ln o - \ln \tau}<\frac{o-\tau+ \ln(Q(\gamma))}{\ln o - \ln \tau}\right]\\
&=\left[X<\frac{o-\tau+ \ln(Q(\gamma))}{\ln o - \ln \tau}\right],
\end{aligned}$
\item 
$\begin{aligned}[t] B_\gamma &=[Y>Q(\gamma)]=
\left[\frac{o-\tau+ \ln(Y(\omega))}{\ln o - \ln \tau}>\frac{o-\tau+ \ln(Q(\gamma))}{\ln o - \ln \tau}\right]\\
&=\left[X>\frac{o-\tau+ \ln(Q(\gamma))}{\ln o - \ln \tau}\right],
\end{aligned}$

\item 
$\begin{aligned}[t] C_\gamma &=[Y=Q(\gamma)]=\left[\frac{o-\tau+ \ln(Y(\omega))}{\ln o - \ln \tau}=\frac{o-\tau+ \ln(Q(\gamma))}{\ln o - \ln \tau}\right]\\
&=\left[X=\frac{o-\tau+ \ln(Q(\gamma))}{\ln o - \ln \tau}\right].
\end{aligned}$

\end{itemize}

Since $X\sim \mbox{Poisson}\left(\tau\right)$ we have that $Q_X(\gamma)\in\mathbb{N}$ and $\mu(\{0,...,Q_X(\gamma)-1\}|\tau)<\gamma$ and $\mu(\{0,...,Q_X(\gamma)\}|\tau)\ge\gamma$, by the definitions of $A_\gamma$ and $B_\gamma$ we have that

\begin{itemize}
    \item $P\left[X<\frac{o-\tau+ \ln(Q(\gamma))}{\ln o - \ln \tau}\right]\le\gamma$, e
    \item $P\left[X>\frac{o-\tau+ \ln(Q(\gamma))}{\ln o - \ln \tau}\right]\le1-\gamma$.
\end{itemize}
Therefore, we can define the sets as $A_\gamma=[X<Q_X(\gamma)]$, $B_\gamma=[X>Q_X(\gamma)]$, and $C_\gamma=[X=Q_X(\gamma)]$.

Using the relationship between the Poisson distribution and the chi-square distribution, we have

\begin{itemize}
    \item $\mu(\{0,...,Q_X(\gamma)-1\}|\tau)=1-\int_0^\tau\frac{x^{\frac{2Q_X(\gamma)}{2}-1}e^{-\frac{x}{2}}}{2^\frac{2Q_X(\gamma)}{2}\Gamma(\frac{2Q_X(\gamma)}{2})}d\lambda(x)<\gamma$, e
    \item $\mu(\{0,...,Q_X(\gamma)\}|\tau)=1-\int_0^\tau\frac{x^{\frac{2Q_X(\gamma)+2}{2}-1}e^{-\frac{x}{2}}}{2^\frac{2Q_X(\gamma)+@}{2}\Gamma(\frac{2Q_X(\gamma)+2}{2})}d\lambda(x)\ge\gamma$.
\end{itemize}

These inequalities are equivalent to

\begin{itemize}
    \item $\tau> \chi^2_{2Q_X(\gamma),1-\gamma}$,
    \item $\tau\le \chi^2_{2Q_X(\gamma)+2,1-\gamma}$.
\end{itemize}

With $\chi^2_{q,\gamma}$ being the quantile function of the chi-square distribution with $q$ degrees of freedom as a function of $\gamma$, since the right-hand side of the inequalities can be written as a function of $\chi^2_{2i,1-\gamma}$, with $i=Q_X(\gamma)$ for the upper part and $i=Q_X(\gamma)+1$ for the lower part, being an increasing function as a function of $i$, we can define $Q_X(\gamma)$ as \\
$Q_X(\gamma)=\sup \{i\in\mathbb{N}|\tau> \chi^2_{2i,1-\gamma}\}$.

In other words, we have that $Q_X(\gamma)=i$ if, and only if, $\chi^2_{2i,1-\gamma}<\tau\le \chi^2_{2i+2,1-\gamma}$. Under this condition, we have the definitions of the sets $A_\gamma=\{0,1,...,i-1\}$, $B_\gamma=\{i+1,i+2,...\}$ and $C_\gamma=\{i\}$, and therefore $\mu(A_\gamma|\tau)=\sum_{j=0}^{i-1}\frac{e^{-\tau}\tau^j}{j!}$ and $\mu(C_\gamma|\tau)=\frac{e^{-\tau}\tau^i}{i!}$. By replacing the value of $i$ with $\omega$, it is possible to write the function $\psi_o(\omega \mid \tau)$ for the case $\tau<o$ present in \eqref{eq:por} given by

\begin{equation}
\label{eq:por}
     \psi_o(\omega \mid \tau) = \left\{
    \begin{array}{rcl}
        1&\mbox{, if}&
        \tau> \chi^2_{2\omega+2,1-\gamma},
        \\
        0&\mbox{, if}&
        \tau\le \chi^2_{2\omega,1-\gamma},
        \\
        \frac{\gamma-\sum_{i=0}^{\omega-1}\frac{e^{-\tau}\tau^i}{i!}}{\frac{e^{-\tau}\tau^\omega}{\omega!}}
        &\mbox{, if}&
        \chi^2_{2\omega,1-\gamma}<\tau\le \chi^2_{2\omega+2,1-\gamma}.
    \end{array} \right.
\end{equation}

\begin{figure}[H]
\graphicspath{{img/}}
\centering
\includegraphics[width=15cm]{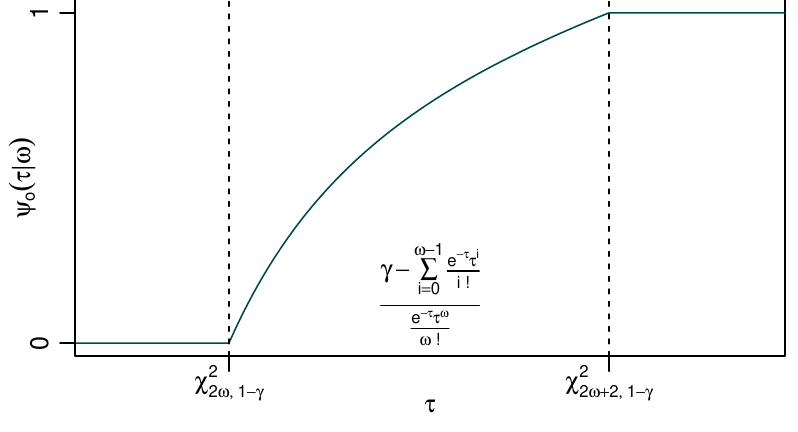}
\caption{Fuzzy membership function of the method developed for the value of $\tau$ for the Poisson case with $\tau<o$.}
\label{fig:poisson1}
\end{figure}

This function is non-decreasing with respect to $\tau$ with $\omega$ fixed, since it is the case $\tau<o$, the fuzzy membership function of the described method is non-decreasing in $\tau$ up to the value $o$. It is shown in Figure \ref{fig:poisson1}.

\begin{figure}[H]
\graphicspath{{img/}}
\includegraphics[width=15cm]{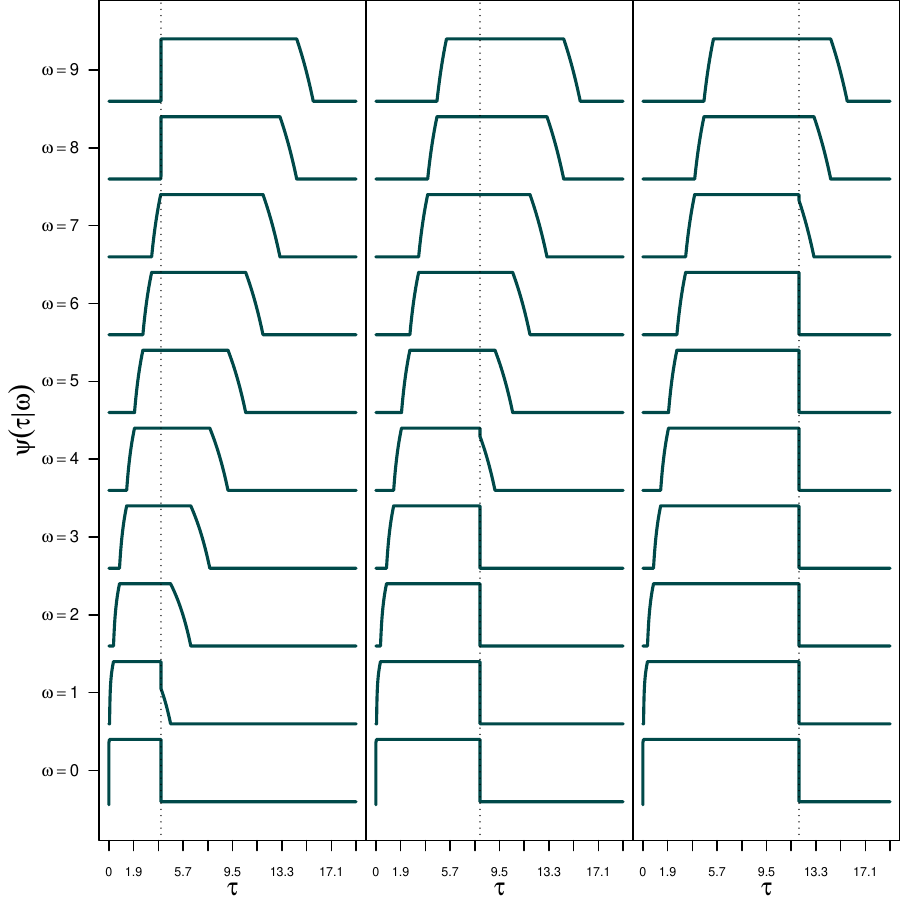}
\caption{Fuzzy membership function $\psi_o(\omega \mid \tau)$ in the case of the Poisson distribution with $o=4$ for the left panel, $o=8$ for the center panel and $o=12$ for the right panel.} \label{fig:poisson2}
\centering
\end{figure}

\begin{figure}[H]
\graphicspath{{img/}}
\includegraphics[width=15cm]{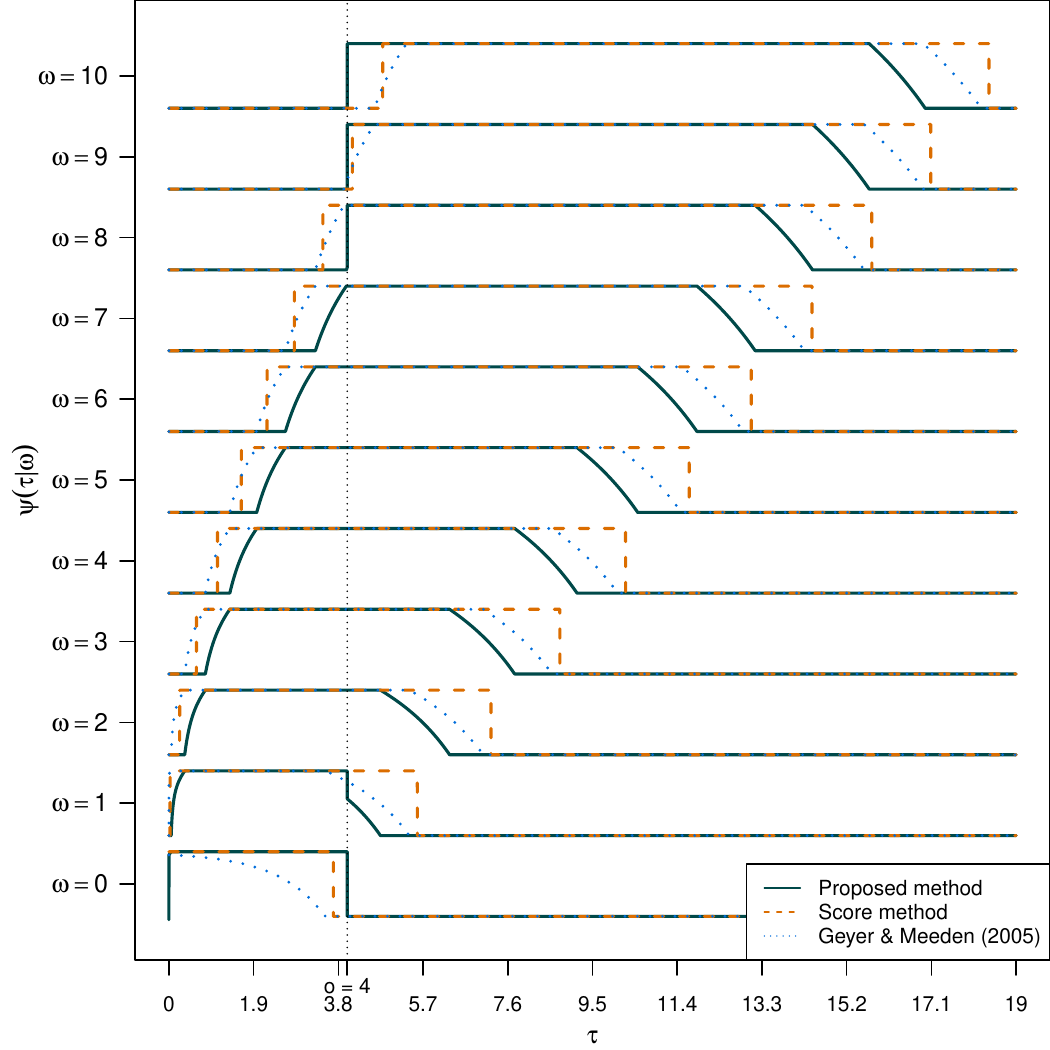}
\caption{Fuzzy membership functions $\psi_{3.8}$, $\psi^{GM}$ and $\psi^{S}$ with confidence level $\gamma = 0.95$ for the Poisson distribution.} \label{fig:poisson3}
\centering
\end{figure}

Similar for $\tau > o$. Combining the two cases, the function $\psi_o$ is given by \eqref{eq:psiopoiss}, the function is non-decreasing up to $o$ and non-increasing after that value when evaluated with respect to $\tau$ with $\omega$ and $o$ fixed, as presented in Figure \ref{fig:poisson2} for $o = 4$, $8$ and $12$. The main characteristic of this function is that its left-hand limit or its right-hand limit at $o$ is equal to $1$, with a discontinuity at $\omega$ far from the value $\lfloor o(n+1)\rfloor$, containing only values less than $o$ if $\omega<\lfloor o(n+1)\rfloor$, and only values greater than $o$ if $\omega>\lfloor o(n+1)\rfloor$. Figure \ref{fig:poisson3} compares the proposed method for $o=3.8$ and the Geyer-Meeden and Score methods when $\gamma = 0.95$.

\begin{equation}
\label{eq:psiopoiss}
     \psi_o(\omega \mid \tau) = \left\{
    \begin{array}{rcl}
        1&\mbox{, if}&
        \tau> \chi^2_{2\omega+2,1-\gamma}\mbox{ and } \tau<o
        \\
        1&\mbox{, if}&
        \tau\le \chi^2_{2\omega,\gamma}\mbox{ and }\tau>o
        \\
        0&\mbox{, if}&
        \tau\le \chi^2_{2\omega,1-\gamma}\mbox{ and }\tau<o
        \\
        0&\mbox{, if}&
        \tau> \chi^2_{2\omega+2,\gamma}\mbox{ and }\tau>o
        \\
        \frac{\gamma-\sum_{i=0}^{\omega-1}\frac{e^{-\tau}\tau^i}{i!}}{\frac{e^{-\tau}\tau^\omega}{\omega!}}
        &\mbox{, if}&
        \chi^2_{2\omega,1-\gamma}<\tau\le \chi^2_{2\omega+2,1-\gamma}\mbox{ and }\tau<o\\
        \frac{\gamma-1+\sum_{i=0}^{\omega}\frac{e^{-\tau}\tau^i}{i!}}{\frac{e^{-\tau}\tau^\omega}{\omega!}}
        &\mbox{, if}&
        \chi^2_{2\omega,\gamma}<\tau\le \chi^2_{2\omega+2,\gamma}\mbox{ and }\tau>o.
    \end{array} \right.
\end{equation}

\subsubsection{Expected interval length}

\begin{figure}[H]
\graphicspath{{img/}}
\includegraphics[width=13cm]{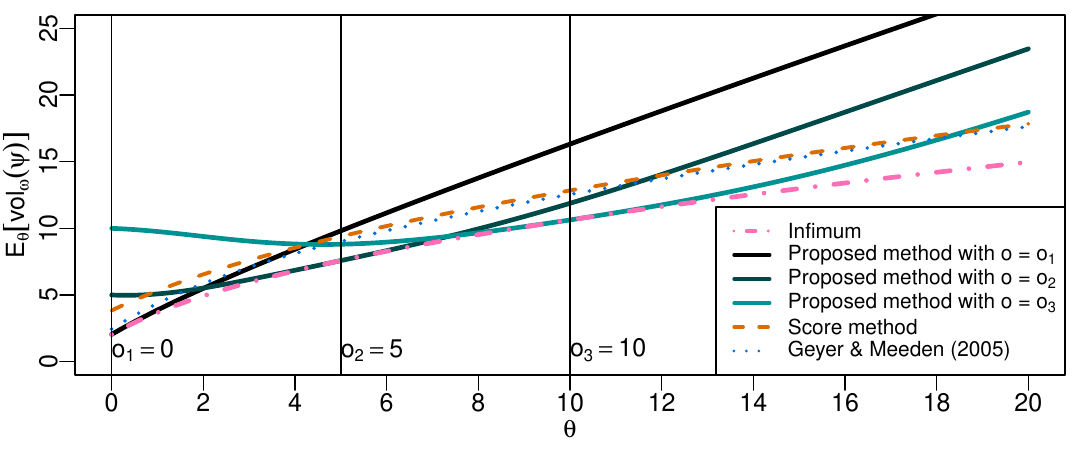}
\caption{Expected interval length in the Poisson case with parameter $\theta$, for the five methods discussed, three of which correspond to $\psi_o$ for $o_1 = 0$, $o_2 = 5$, and $o_3 = 10$, and the methods $\psi^{GM}$ and $\psi^{S}$. The black dotted curve represents the lower bound obtained by $TE(\theta, \psi_\theta, \lambda)$, where $\lambda$ refers to the Lebesgue measure.} \label{fig:poisson4}
\centering
\end{figure}

In this example, we compute the expected lengths of the confidence interval for the mean $\theta$ of the Poisson distribution using the proposed method and the methods of Geyer–Meeden and the Score method. As in the case of the binomial distribution, the lengths were calculated numerically and are presented in Figure \ref{fig:poisson4}. From the figure, we observe that the proposed method is tangent to the lower bound at $\theta = o$, as in the other examples. Moreover, the method shows better performance for values of $\theta$ close to $o$ compared to the usual methods. Specifically, for $o = 5$, the expected length obtained by the proposed method is close to the lengths computed by the other methods for values of $\theta$ near zero.

\subsection{Normal distribution}

\subsubsection{Fuzzy pertinent function}
Consider the case of $n$ independently distributed observations with a normal distribution with mean $\theta\in\Theta=\mathbb{R}$ and known standard deviation $\sigma>0$. In this case, $\Omega\in\mathbb{R}^n ,\ \mathcal{A}=\mathcal{B}(\mathbb{R}^n)$, the Lebesgue measure $\lambda:\mathcal{B}(\mathbb{R}^n)\to[0,\infty]$ and 
$$\displaystyle \mu(A|\theta)=\int_A(2\pi\sigma^2)^{-\frac{n}{2}}\exp(-\frac{1}{2\sigma^2}(\omega-\theta\mathbf{1})^t(\omega-\theta\mathbf{1}))d\lambda(\omega).$$

In this case, $D^c=\mathbb{R}^n.$ Furthermore, 
\begin{eqnarray*}
Y(\omega)&=&\frac{d\mu(\cdot \mid o)}{d\mu(\cdot \mid \tau)}(\omega) \\
&=& \exp\left(-\frac{1}{2\sigma^2}[(\omega-o1)^t(\omega-o1)-(\omega-\tau1)^t(\omega-\tau1)]\right)\\ &=& \exp\left(\frac{1}{\sigma^2}(o-\tau)\omega^t1\right)\exp\left(-\frac{n}{2\sigma^2}(o^2-\tau^2)\right).
\end{eqnarray*}

To define the sets $A_\gamma$, $B_\gamma$, and $C_\gamma$, it is possible to calculate the quantile function of $Y$ in $\gamma$ or $Q(\gamma)$, but it is easier to find an equivalence to the sets $[Y<Q(\gamma)]$, $[Y=Q(\gamma)]$, and $[Y>Q(\gamma)]$ using an auxiliary random variable.

Defining the random variable by the function $X(\omega)=\frac{\omega^t1}{n}$, we have $X\sim N\left(\tau,\frac{\sigma^2}{n}\right)$ and $Y(\omega)=\exp\left(\frac{1}{\sigma^2}(o-\tau)nX(\omega)\right)\exp\left(-\frac{n}{2\sigma^2}(o^2-\tau^2)\right)$, for $o\neq \tau$, we can rewrite it as $X(\omega)=\frac{\sigma^2}{n(o-\tau)}\ln (Y(\omega))+\frac{o+\tau}{2}$, note that the function is increasing in $X(\omega)$ for $o>\tau$ and decreasing if If $o<\tau$, then we will divide it into two cases and write the sets $A_\gamma$, $B_\gamma$, $C_\gamma$ as a function of $X$.
\paragraph*{Case 1: $o>\tau$}

\begin{itemize}
\item 
$\begin{aligned}[t] A_\gamma &=[Y<Q(\gamma)]=\left[\frac{\sigma^2}{n(o-\tau)}\ln (Y)+\frac{o+\tau}{2}<\frac{\sigma^2}{n(o-\tau)}\ln (Q(\gamma))+\frac{o+\tau}{2}\right]\\
&=\left[X<\frac{\sigma^2}{n(o-\tau)}\ln (Q(\gamma))+\frac{o+\tau}{2}\right],
\end{aligned}$
\item 
$\begin{aligned}[t] B_\gamma &=[Y>Q(\gamma)]=\left[\frac{\sigma^2}{n(o-\tau)}\ln (Y)+\frac{o+\tau}{2}>\frac{\sigma^2}{n(o-\tau)}\ln (Q(\gamma))+\frac{o+\tau}{2}\right]\\
&=\left[X>\frac{\sigma^2}{n(o-\tau)}\ln (Q(\gamma))+\frac{o+\tau}{2}\right],
\end{aligned}$

\item 
$\begin{aligned}[t] C_\gamma &=[Y=Q(\gamma)]=\left[\frac{\sigma^2}{n(o-\tau)}\ln (Y)+\frac{o+\tau}{2}=\frac{\sigma^2}{n(o-\tau)}\ln (Q(\gamma))+\frac{o+\tau}{2}\right]\\
&=\left[X=\frac{\sigma^2}{n(o-\tau)}\ln (Q(\gamma))+\frac{o+\tau}{2}\right].
\end{aligned}$

\end{itemize}

Note that these definitions imply that $P[C_\gamma]=0$, since $X$ has a Normal distribution and the cumulative probability at a point is equal to 0, therefore $P[A_\gamma]=\gamma$ and $P[B_\gamma]=1-\gamma$. Furthermore, the quantile function of $X$ can be written as a function of the value $\frac{\sigma^2}{n(o-\tau)}\ln (Q(\gamma))+\frac{o+\tau}{2}$ and since $P\left[X<\frac{\sigma^2}{n(o-\tau)}\ln (Q(\gamma))+\frac{o+\tau}{2}\right]=\gamma$ then $Q_X(\gamma)=\frac{\sigma^2}{n(o-\tau)}\ln (Q(\gamma))+\frac{o+\tau}{2}$.

Since $X\sim N\left(\tau,\frac{\sigma^2}{n}\right)$ we have that $Q_X(\gamma)=\tau+Z_\gamma\frac{\sigma}{\sqrt n}$ and therefore

\begin{itemize}
    \item $A_\gamma=
    \left[X<\tau+Z_\gamma\frac{\sigma}{\sqrt n}\right]=
    \left[\tau>X-Z_\gamma\frac{\sigma}{\sqrt n}\right],$
    \item $B_\gamma=
    \left[X>\tau+Z_\gamma\frac{\sigma}{\sqrt n}\right]=
    \left[\tau<X-Z_\gamma\frac{\sigma}{\sqrt n}\right],$
    \item $C_\gamma=
    \left[X=\tau+Z_\gamma\frac{\sigma}{\sqrt n}\right]=
    \left[\tau=X-Z_\gamma\frac{\sigma}{\sqrt n}\right].$
\end{itemize}

Then, for $o>\tau$ 

$$\psi_\theta(x,\tau)=\left\{
\begin{array}{rcl}
  1 & , \mbox{ if } &\tau>x-Z_\gamma\frac{\sigma}{\sqrt n}\ \ and\ \ o>\tau, \\
  0 & , \mbox{ if } & \tau \le x-Z_\gamma\frac{\sigma}{\sqrt n}\ \ and\ \ o>\tau.
\end{array}
\right.$$

Similarly, for $o<\tau$ 

$$\psi_o(x,\tau)=\left\{
\begin{array}{rcl}
  1 & , \mbox{ if } &\tau<x+Z_\gamma\frac{\sigma}{\sqrt n}\ \ and\ \ o<\tau, \\
  0 & , \mbox{ if } & \tau \ge x+Z_\gamma\frac{\sigma}{\sqrt n}\ \ and\ \ o<\tau.
\end{array}
\right.$$

Combining the two cases, we obtain the following fuzzy membership function for the described method.
\begin{eqnarray*}
\psi_o(x,\tau)&=&\left\{\begin{array}{rll}
  1 & , \mbox{ if } &\tau>x-Z_\gamma\frac{\sigma}{\sqrt n}\ \ e\ \ o>\tau \\
  1 & , \mbox{ if } &\tau<x+Z_\gamma\frac{\sigma}{\sqrt n}\ \ e\ \ o<\tau \\
  0 & ,& \mbox{ otherwise.}
\end{array}\right.  \\
&=& \left\{\begin{array}{ll}
1 &, \mbox{ if }  \tau\in\left(\min\left(o,x-Z_\gamma\sqrt\frac{\sigma^2}{n}\right),\max\left(o,x+Z_\gamma\sqrt\frac{\sigma^2}{n}\right)\right)\\
0 &, \mbox{ otherwise.}
\end{array}\right.
\end{eqnarray*}

\begin{figure}[H]
\graphicspath{{img/}}
\includegraphics[width=15cm]{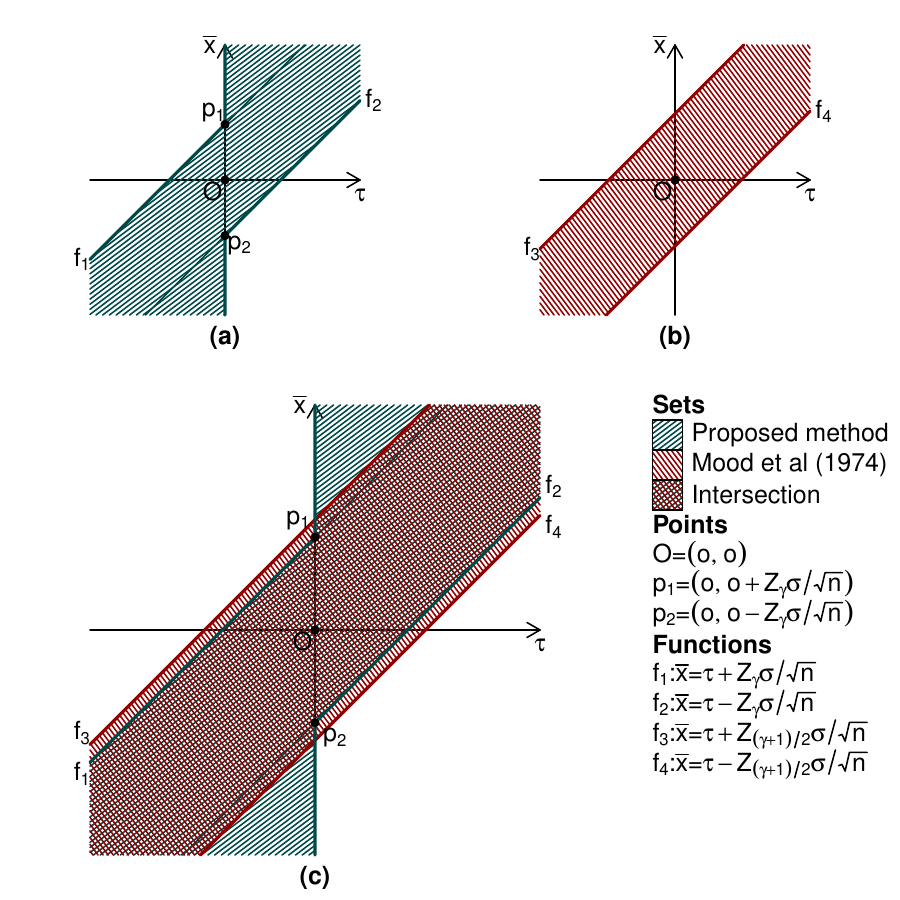}
\caption{Functions $\psi_0$ (upper left panel) and $\psi^N$ (upper right panel). The shaded region indicates the set of values for which the function equals $1$, while in the white region the function equals $0$. The dotted lines represent the boundaries of this region. The lower panel shows the overlap of these regions.} \label{fig:normal1}
\centering
\end{figure}

To compare the results, we will use the commonly used confidence interval. In both cases, the functions can be written as a function of the statistic $x=\frac{w^t1}{n}$ and $\tau$, and only assume the values $0$ or $1$, and therefore both fit the classic case of confidence intervals. Because of this, it is possible to visualize $\psi_o$ and $\psi^*$ in a graph with the axes representing $\frac{w^t1}{n}$ and $\tau$ and the region where the value $1$ is assumed, as shown in Figure \ref{fig:normal1}.

Consider the previous example, making the following modification to the parameter space $\Theta=[a,b] \subset \mathbb{R}$. With this modification, the new membership function for the standard method $\psi^{NL}$ has the same value as $\psi^N$ for $\tau\in[a,b]$.

\begin{figure}[H]
\graphicspath{{img/}}
\includegraphics[width=15cm]{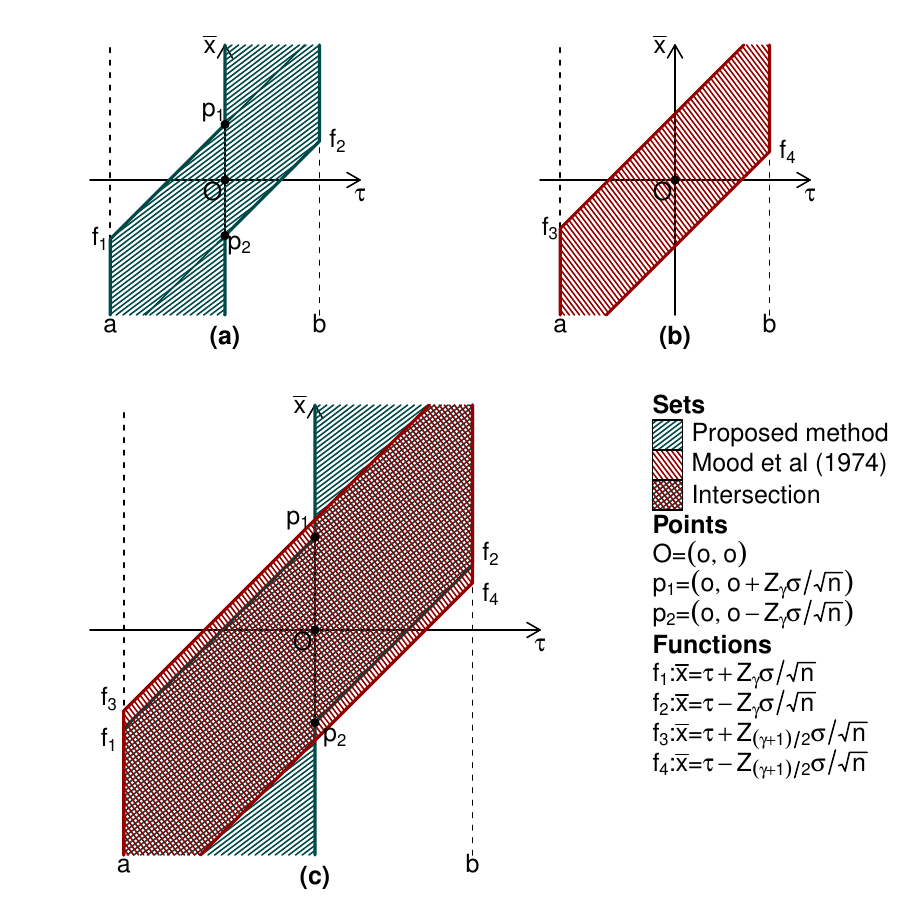}
\caption{Functions $\psi_0$ (left panel) and $\psi^N$ (right panel). The blue region indicates where the function value equals $1$, while outside this region the value is $0$. The dotted lines define the boundary of this region.} \label{fig:normal2}
\centering
\end{figure}

In this case, Figure \ref{fig:normal2} shows the confidence region for the case where the parameter space is bounded.

\subsubsection{Expected interval length}
In this example, we compute the expected length of the confidence interval for the mean $\theta$ of the normal distribution in the case $\theta \in (a,b)$, $a < b$, that is, when the mean is known to be bounded. In both cases, we first compute $\int_\Theta \psi(\omega \mid  \tau)\,d\lambda(\tau)$ by splitting into regions, and then integrate it with respect to the measure $\mu(\cdot|\theta)$.

For $\psi_o$ we have

\begin{eqnarray*}
    \lefteqn{\int_\Theta \psi_o(\omega \mid \tau)\,d\lambda(\tau) =} \\
    && 
    \quad 1_{(b-Z_\gamma\frac{\sigma}{\sqrt{n}},\infty)}(\omega)(b-o)
    +1_{(a+Z_\gamma\frac{\sigma}{\sqrt{n}},o+Z_\gamma\frac{\sigma}{\sqrt{n}})}(\omega)(o+Z_\gamma\frac{\sigma}{\sqrt{n}}-\omega) \\
    && \quad \quad + 1_{(-\infty,a+Z_\gamma\frac{\sigma}{\sqrt{n}})}(\omega)(o-a) 
    + 1_{(o-Z_\gamma\frac{\sigma}{\sqrt{n}},b-Z_\gamma\frac{\sigma}{\sqrt{n}})}(\omega)(\omega-(o-Z_\gamma\frac{\sigma}{\sqrt{n}})).
\end{eqnarray*}

Using the identities
\[
\int_{-\infty}^z (2\pi)^{-1/2}\exp\!\left(-\frac{x^2}{2}\right)dx = \Phi(z), \]
and
\[\int_{z_1}^{z_2} x\exp\!\left(-\frac{x^2}{2}\right)dx = \exp\!\left(-\frac{z_1^2}{2}\right) - \exp\!\left(-\frac{z_2^2}{2}\right),
\]
the value of $TE(\psi_o,\theta,\lambda)$ is obtained from the following expression:
\[
\begin{array}{rcl}
     \mathrm{EL}(\psi_o,\theta,\lambda) &=& 
     (b-o)\bigl(1-\Phi(\frac{\sqrt{n}}{\sigma}(b-Z_\gamma\frac{\sigma}{\sqrt{n}}-\theta))\bigr) \\
     &+& (o-a)\Phi(\frac{\sqrt{n}}{\sigma}(a+Z_\gamma\frac{\sigma}{\sqrt{n}}-\theta))  \\
     &+& (\theta-(o-Z_\gamma\frac{\sigma}{\sqrt{n}}))\bigl(\Phi(\frac{\sqrt{n}}{\sigma}(b-Z_\gamma\frac{\sigma}{\sqrt{n}}-\theta))-\Phi(\frac{\sqrt{n}}{\sigma}(o-Z_\gamma\frac{\sigma}{\sqrt{n}}-\theta))\bigr) \\
     &+& \frac{\sigma}{\sqrt{2\pi n}}\bigl(\exp(-\frac{n}{2 \sigma^2}(o-Z_\gamma\frac{\sigma}{\sqrt{n}}-\theta)^2)-\exp(-\frac{n}{2 \sigma^2}(b-Z_\gamma\frac{\sigma}{\sqrt{n}}-\theta)^2)\bigr)  \\
     &+& (o+Z_\gamma\frac{\sigma}{\sqrt{n}}-\theta)\bigl(\Phi(\frac{\sqrt{n}}{\sigma}(o+Z_\gamma\frac{\sigma}{\sqrt{n}}-\theta))-\Phi(\frac{\sqrt{n}}{\sigma}(a+Z_\gamma\frac{\sigma}{\sqrt{n}}-\theta))\bigr)\\
     &-& \frac{\sigma}{\sqrt{2\pi n}}\bigl(\exp(-\frac{n}{2 \sigma^2}(a+Z_\gamma\frac{\sigma}{\sqrt{n}}-\theta)^2)-\exp(-\frac{n}{2 \sigma^2}(o+Z_\gamma\frac{\sigma}{\sqrt{n}}-\theta)^2)\bigr).
\end{array}
\]

For $\psi^{NL}$ we must divide into two cases, corresponding to the conditions 
$a+Z_{\frac{1+\gamma}{2}}\frac{\sigma}{\sqrt{n}} < b-Z_{\frac{1+\gamma}{2}}\frac{\sigma}{\sqrt{n}}$ 
and 
$a+Z_{\frac{1+\gamma}{2}}\frac{\sigma}{\sqrt{n}} > b-Z_{\frac{1+\gamma}{2}}\frac{\sigma}{\sqrt{n}}$.

\paragraph*{Case 1:} If $a+Z_{\frac{1+\gamma}{2}}\frac{\sigma}{\sqrt{n}} < b-Z_{\frac{1+\gamma}{2}}\frac{\sigma}{\sqrt{n}}$, then
\[
\begin{array}{rcl}
     \int_\Theta\psi^{NL}(\omega \mid \tau)\, d\lambda(\tau) &=&
     1_{(a-Z_{\frac{1+\gamma}{2}}\frac{\sigma}{\sqrt{n}},\,a+Z_{\frac{1+\gamma}{2}}\frac{\sigma}{\sqrt{n}})}(\omega)(\omega-a+Z_{\frac{1+\gamma}{2}}\frac{\sigma}{\sqrt{n}}) \\
     &+& 1_{(a+Z_{\frac{1+\gamma}{2}}\frac{\sigma}{\sqrt{n}},\,b-Z_{\frac{1+\gamma}{2}}\frac{\sigma}{\sqrt{n}})}(\omega)\,2Z_{\frac{1+\gamma}{2}}\frac{\sigma}{\sqrt{n}} \\
     &+& 1_{(b-Z_{\frac{1+\gamma}{2}}\frac{\sigma}{\sqrt{n}},\,b+Z_{\frac{1+\gamma}{2}}\frac{\sigma}{\sqrt{n}})}(\omega)(b+Z_{\frac{1+\gamma}{2}}\frac{\sigma}{\sqrt{n}}-\omega).
\end{array}
\]

Thus,
\begin{eqnarray*}
    \lefteqn{ \mathrm{EL}(\psi^{NL},\theta,\lambda) =} \\
    &=& (\theta-a+Z_{\frac{1+\gamma}{2}}\frac{\sigma}{\sqrt{n}})
     \bigl(\Phi(\frac{\sqrt{n}}{\sigma}(a-\theta)+Z_{\frac{1+\gamma}{2}})-\Phi(\frac{\sqrt{n}}{\sigma}(a-\theta)-Z_{\frac{1+\gamma}{2}})\bigr)
     \\
     &+& \frac{\sigma}{\sqrt{2\pi n}}\Bigl(
     \exp(-\frac{n}{2\sigma^2}(a-Z_{\frac{1+\gamma}{2}}\frac{\sigma}{\sqrt{n}}-\theta)^2)
     -\exp(-\frac{n}{2\sigma^2}(a+Z_{\frac{1+\gamma}{2}}\frac{\sigma}{\sqrt{n}}-\theta)^2)
     \Bigr) \\
     &+& 2Z_{\frac{1+\gamma}{2}}\frac{\sigma}{\sqrt{n}}
     \bigl(\Phi(\frac{\sqrt{n}}{\sigma}(b-\theta)+Z_{\frac{1+\gamma}{2}})-\Phi(\frac{\sqrt{n}}{\sigma}(a-\theta)-Z_{\frac{1+\gamma}{2}})\bigr) \\
     &+& (b+Z_{\frac{1+\gamma}{2}}\frac{\sigma}{\sqrt{n}}-\theta)
     \bigl(\Phi(\frac{\sqrt{n}}{\sigma}(b-\theta)+Z_{\frac{1+\gamma}{2}})-\Phi(\frac{\sqrt{n}}{\sigma}(b-\theta)-Z_{\frac{1+\gamma}{2}})\bigr) \\
     &-& \frac{\sigma}{\sqrt{2\pi n}}\Bigl(
     \exp(-\frac{n}{2\sigma^2}(b-Z_{\frac{1+\gamma}{2}}\frac{\sigma}{\sqrt{n}}-\theta)^2)
     -\exp(-\frac{n}{2\sigma^2}(b+Z_{\frac{1+\gamma}{2}}\frac{\sigma}{\sqrt{n}}-\theta)^2)
     \Bigr).
\end{eqnarray*}

\paragraph*{Case 2:} If $a+Z_{\frac{1+\gamma}{2}}\frac{\sigma}{\sqrt{n}} > b-Z_{\frac{1+\gamma}{2}}\frac{\sigma}{\sqrt{n}}$, then
\[
\begin{array}{rcl}
     \int_\Theta\psi^{NL}(\omega \mid \tau)\, d\lambda(\tau) &=&
     1_{(a-Z_{\frac{1+\gamma}{2}}\frac{\sigma}{\sqrt{n}},\,b-Z_{\frac{1+\gamma}{2}}\frac{\sigma}{\sqrt{n}})}(\omega)(\omega-a+Z_{\frac{1+\gamma}{2}}\frac{\sigma}{\sqrt{n}}) \\
     &+& 1_{(b-Z_{\frac{1+\gamma}{2}}\frac{\sigma}{\sqrt{n}},\,a+Z_{\frac{1+\gamma}{2}}\frac{\sigma}{\sqrt{n}})}(\omega)(b-a) \\
     &+& 1_{(a+Z_{\frac{1+\gamma}{2}}\frac{\sigma}{\sqrt{n}},\,b+Z_{\frac{1+\gamma}{2}}\frac{\sigma}{\sqrt{n}})}(\omega)(b+Z_{\frac{1+\gamma}{2}}\frac{\sigma}{\sqrt{n}}-\omega).
\end{array}
\]

Thus,
\begin{eqnarray*}
    \lefteqn{\mathrm{EL}(\psi^{NL},\theta,\lambda) =} \\
    &&     (\theta-a+Z_{\frac{1+\gamma}{2}}\frac{\sigma}{\sqrt{n}})
     \bigl(\Phi(\frac{\sqrt{n}}{\sigma}(b-\theta)-Z_{\frac{1+\gamma}{2}})-\Phi(\frac{\sqrt{n}}{\sigma}(a-\theta)-Z_{\frac{1+\gamma}{2}})\bigr)
     \\
     &+& \frac{\sigma}{\sqrt{2\pi n}}
     \exp(-\frac{n}{2\sigma^2}(a-Z_{\frac{1+\gamma}{2}}\frac{\sigma}{\sqrt{n}}-\theta)^2)
     -\frac{\sigma}{\sqrt{2\pi n}}
     \exp(-\frac{n}{2\sigma^2}(b-Z_{\frac{1+\gamma}{2}}\frac{\sigma}{\sqrt{n}}-\theta)^2) \\
     &+& (b-a)
     \bigl(\Phi(\frac{\sqrt{n}}{\sigma}(a-\theta)+Z_{\frac{1+\gamma}{2}})-\Phi(\frac{\sqrt{n}}{\sigma}(b-\theta)-Z_{\frac{1+\gamma}{2}})\bigr) \\
     &+& (b+Z_{\frac{1+\gamma}{2}}\frac{\sigma}{\sqrt{n}}-\theta)
     \bigl(\Phi(\frac{\sqrt{n}}{\sigma}(b-\theta)+Z_{\frac{1+\gamma}{2}})-\Phi(\frac{\sqrt{n}}{\sigma}(a-\theta)+Z_{\frac{1+\gamma}{2}})\bigr) \\
     &-& \frac{\sigma}{\sqrt{2\pi n}}
     \exp(-\frac{n}{2\sigma^2}(a+Z_{\frac{1+\gamma}{2}}\frac{\sigma}{\sqrt{n}}-\theta)^2)
     + \frac{\sigma}{\sqrt{2\pi n}}
     \exp(-\frac{n}{2\sigma^2}(b+Z_{\frac{1+\gamma}{2}}\frac{\sigma}{\sqrt{n}}-\theta)^2).
\end{eqnarray*}

The lower bound is given by
\[
\begin{array}{rcl}
     \mathrm{EL}(\psi_\theta,\theta,\lambda) &=&
     (b-\theta)(1-\Phi(\frac{\sqrt{n}}{\sigma}(b-Z_\gamma\frac{\sigma}{\sqrt{n}}-\theta))) \\
     &+& (\theta-a)\Phi(\frac{\sqrt{n}}{\sigma}(a+Z_\gamma\frac{\sigma}{\sqrt{n}}-\theta))  \\
     &+& Z_\gamma\frac{\sigma}{\sqrt{n}}\bigl(\Phi(\frac{\sqrt{n}}{\sigma}(b-Z_\gamma\frac{\sigma}{\sqrt{n}}-\theta))-1+\gamma\bigr) \\
     &+& \frac{\sigma}{\sqrt{2\pi n}}\bigl(\exp(-\frac{Z_\gamma^2}{2})-\exp(-\frac{n}{2 \sigma^2}(b-Z_\gamma\frac{\sigma}{\sqrt{n}}-\theta)^2)\bigr)  \\
     &+& Z_\gamma\frac{\sigma}{\sqrt{n}}\bigl(\gamma-\Phi(\frac{\sqrt{n}}{\sigma}(a+Z_\gamma\frac{\sigma}{\sqrt{n}}-\theta))\bigr)\\
     &-& \frac{\sigma}{\sqrt{2\pi n}}\bigl(\exp(-\frac{n}{2 \sigma^2}(a+Z_\gamma\frac{\sigma}{\sqrt{n}}-\theta)^2)-\exp(-\frac{Z_\gamma^2}{2})\bigr).
\end{array}
\]

\begin{figure}[H]
\graphicspath{{img/}}
  \centering{
    \includegraphics[width=0.35\linewidth]{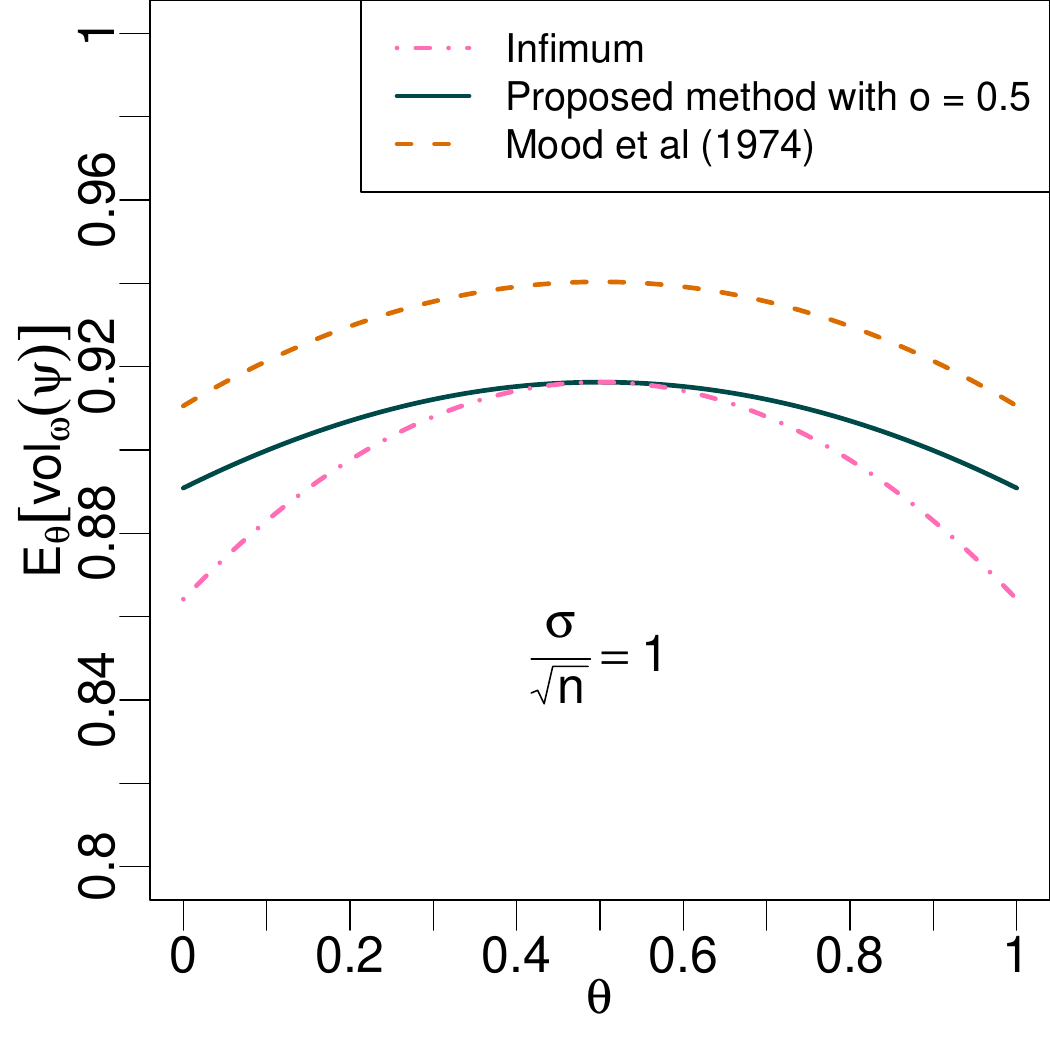}}
    {
    \includegraphics[width=0.35\linewidth]{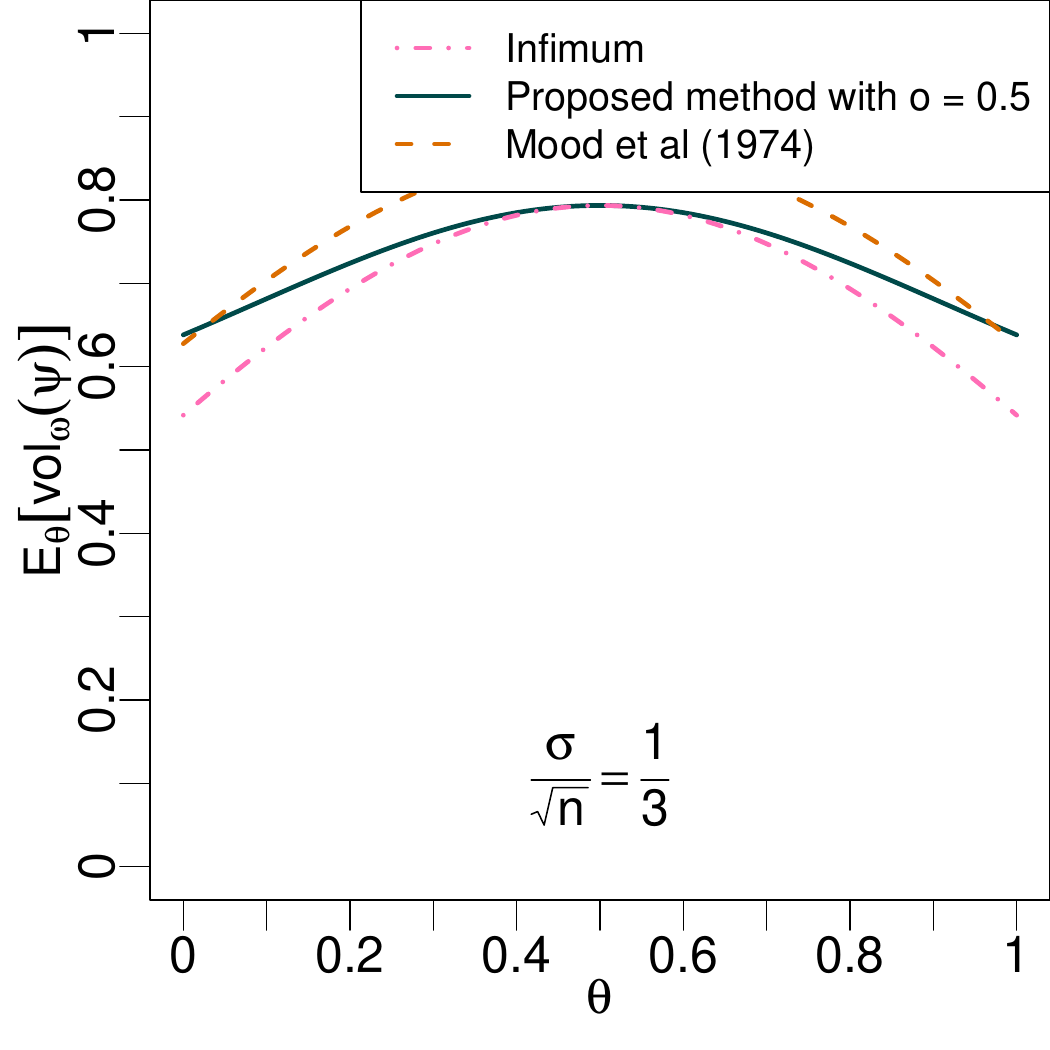}}
  {%
    \includegraphics[width=0.35\linewidth]{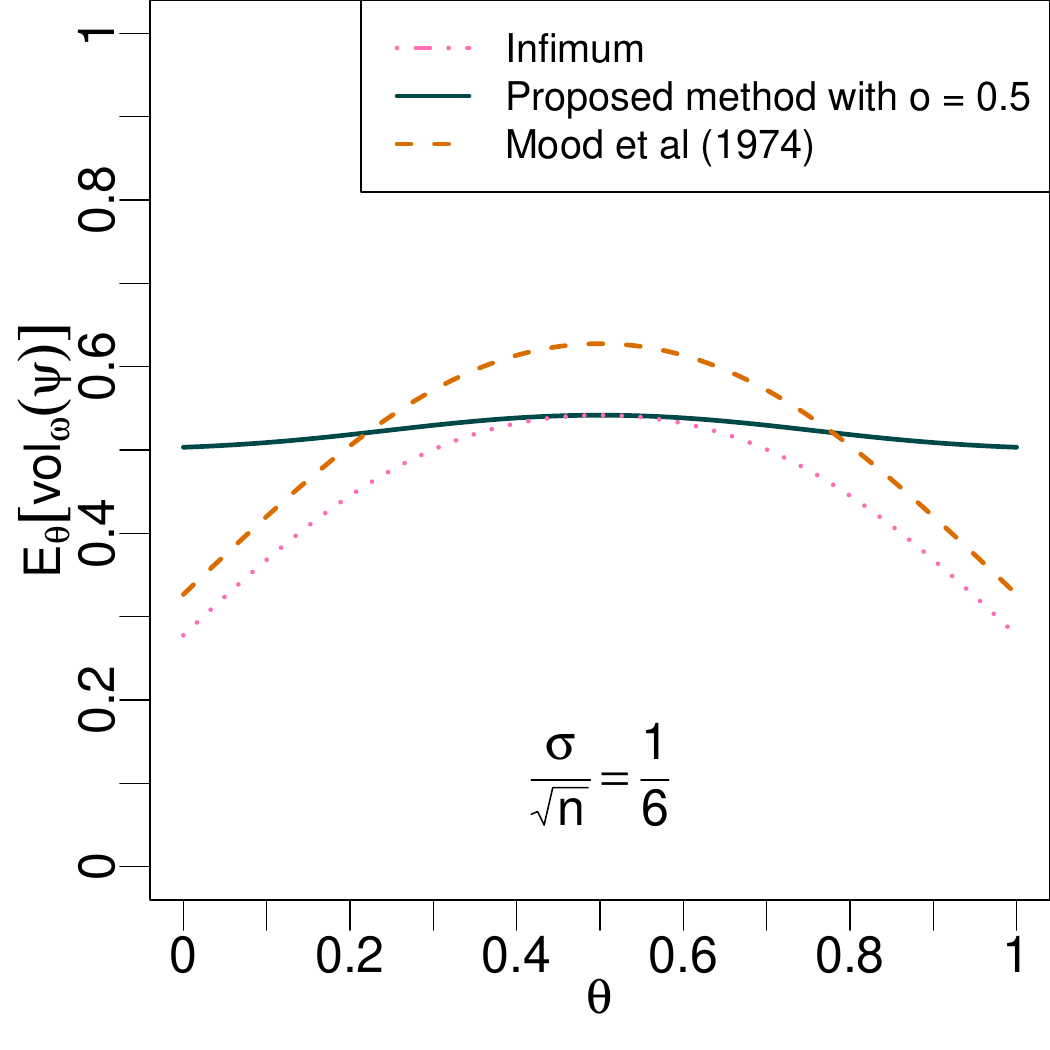}}
  {%
    \includegraphics[width=0.35\linewidth]{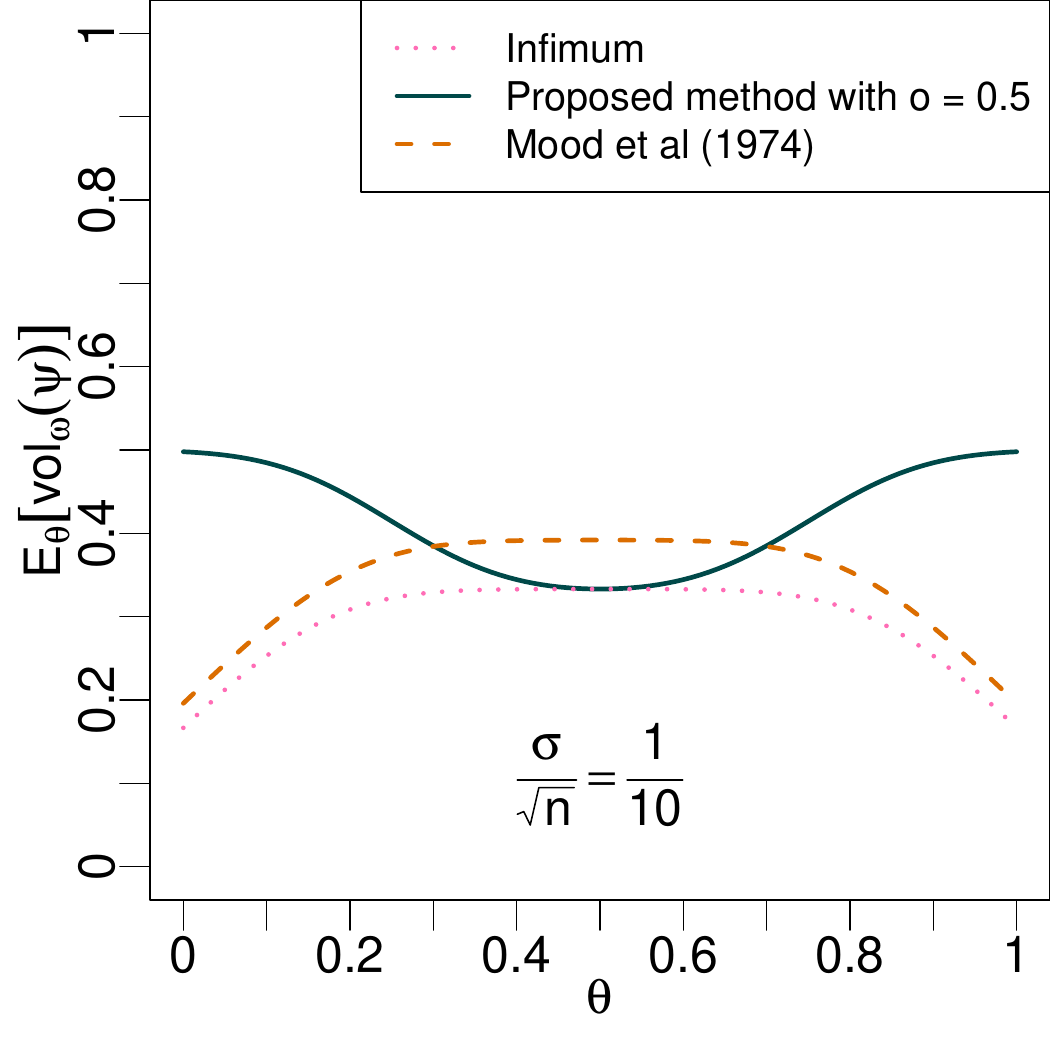}}
    \caption{Expected length in the case of the normal distribution with mean parameter $\mu\in[0,1]$ and variance $\sigma^2$, for $\psi^{NL}$ and $\psi_o$, with $o=0.5$. The black dotted curve shows the lower bound obtained from $\mathrm{EL}(\theta,\psi_\theta,\lambda)$, where $\lambda$ denotes the Lebesgue measure. Each panel uses a different value of $\frac{\sigma}{\sqrt{n}}$.}
    \label{fig:normal3}
\end{figure}

Figure \ref{fig:normal3} shows the expected lengths computed using the proposed method and the method available in the literature for different values of the standard error $\frac{\sigma}{\sqrt{n}}$, namely $\frac{\sigma}{\sqrt{n}} \in \{1/10, 1/6, 1/3, 1\}$. It is possible to observe differences between the behavior of the expected lengths for the different standard errors considered. Furthermore, the proposed method exhibits better performance compared to the method from the literature for larger values of the standard error, that is, it yields smaller expected lengths. It can also be seen that as the standard error decreases, the behavior of the proposed method approaches the behavior obtained for the unrestricted mean case presented in the previous example.

\section{Final Considerations}

The main objective of this work was to propose a new method for constructing fuzzy confidence intervals based on the Neyman-Pearson lemma for simple hypotheses.

The performance of the proposed method was evaluated for constructing randomized confidence intervals for the mean of the normal distribution, the proportion of the binomial distribution, and the mean of the Poisson distribution, and it was compared with standard methods for obtaining such intervals.

For the discrete distributions (binomial and Poisson), the proposed method showed superior performance in terms of the expected interval length, particularly in scenarios where the literature indicates that the standard methods are not suitable for application,  while maintaining appropriate coverage probabilities.

For the the normal distribution, the proposed method performed better when the mean was bounded and the variance was large relative to the range of the mean. As expected, in the case where the mean is unbounded, the method did not show competitive performance compared to the standard one.

A key strength of the proposed method is its broad generality since it is not limited to random variables but provides a unified framework applicable to a wide range of inferential settings. This level of generality significantly extends the scope of confidence interval construction and opens the door to novel applications, such as point processes and random fields which will be investigated in future work.

\section{Acknowledgments}
This study was financed in part by the Coordenação de Aperfeiçoamento de Pessoal de Nível Superior – Brasil (CAPES) – Finance Code 001.  N.L.G. was financed by FAPESP grants 2019/04535-2 and 2023/13453-5, and CNPq grants 304148/2020-2 and 306496/2024-0.

\bibliographystyle{abbrvnat}
\bibliography{references}

\end{document}